\pdfoutput=1
\documentclass[twocolumn,english,aps,superscriptaddress,pre,nofootinbib]{revtex4-2}

\usepackage{babel}
\usepackage{amsmath}
\usepackage{amssymb}
\usepackage{graphicx}
\usepackage{xcolor}
\usepackage{tikz}
\usepackage{hyperref}%
\hypersetup{colorlinks,
citecolor=blue,
filecolor=blue,
linkcolor=blue,
urlcolor=blue,
breaklinks=true
}

\definecolor{bleu}{rgb}{0.29,0.56,0.89}
\definecolor{orange}{rgb}{0.96,0.65,0.14}
\definecolor{rouge}{rgb}{0.82,0.01,0.11}

\begin{document}

\title{Critical line of the triangular Ising antiferromagnet in a field from a $C_3$-symmetric corner transfer matrix algorithm}

\author{Samuel Nyckees}
\email{samuel.nyckees@epfl.ch}
\affiliation{Institute of Physics, Ecole Polytechnique F\'ed\'erale de Lausanne (EPFL), CH-1015 Lausanne, Switzerland}
\author{Afonso Rufino}
\affiliation{Institute of Physics, Ecole Polytechnique F\'ed\'erale de Lausanne (EPFL), CH-1015 Lausanne, Switzerland}

\author{Fr\'ed\'eric Mila}
\affiliation{Institute of Physics, Ecole Polytechnique F\'ed\'erale de Lausanne (EPFL), CH-1015 Lausanne, Switzerland}

\author{Jeanne Colbois}
\email{colbois@irsamc.ups-tlse.fr}
\affiliation{Laboratoire de Physique Th\'eorique, Universit\'e de Toulouse, CNRS, UPS, 31400 Toulouse, France}

\date{\today}
\begin{abstract} 
The corner transfer matrix renormalization group (CTMRG) algorithm has been extensively used to investigate both classical and quantum two-dimensional (2D) lattice models. The convergence of the algorithm can strongly vary from model to model depending on the underlying geometry and symmetries, and the presence of algebraic correlations. An important factor in the convergence of the algorithm is the lattice symmetry, which can be broken due to the necessity of mapping the problem onto the square lattice. We propose a variant of the CTMRG algorithm, designed for models with $C_3$-symmetry, which we apply to the conceptually simple yet numerically challenging problem of the triangular lattice Ising antiferromagnet in a field, at zero and low temperatures. We study how the finite-temperature three-state Potts critical line in this model approaches the ground-state Kosterlitz-Thouless transition driven by a reduced field ($h/T$). In this particular instance, we show that the $C_3$-symmetric CTMRG leads to much more precise results than both existing results from exact diagonalization of transfer matrices and Monte Carlo.
\end{abstract}

\maketitle


\section{Introduction}

\par A very successful approach to study classical spin models on two dimensional lattices is to write the partition function of the system as the contraction of a tensor network~\cite{Nishino1995,Nishino1996,Nishino1997,Levin2007, Orus2009}. Inspired by the original transfer matrix formulation~\cite{KramersWannier1941,Onsager1944}, this tensor network  associates to each interaction a tensor carrying its Boltzmann weight and is typically defined on the same lattice as the original model. With the exception of a few well-known exactly solvable cases, namely zero-field Ising models on planar lattices~\cite{Onsager1944,Kaufman1949,Syozi1951,Wannier1950,Houtappel1950,Fisher1966}, an exact evaluation of this partition function is generally exponentially hard. In those cases, an approximate contraction scheme, such as tensor network renormalization group (TRG, TNR)~\cite{Levin2007,Evenbly2015}, boundary matrix product state (MPS) with infinite time-evolved block decimation (iTEBD)~\cite{Vidal2007,Jordan2008,Orus2008}, variational uniform MPS (VUMPS)~\cite{ZaunerStauber2018,Fishman2018,Nietner2020}, or the corner transfer matrix renormalization group (CTMRG)~\cite{Nishino1995,Nishino1996,Orus2009,Fishman2018} can provide very accurate results. 

\par With the notable exception of TRG/TNR~\cite{Levin2007, Xie2009, Zhao2010, Okunishi2022}, these algorithms are, in most cases, formulated on the square lattice. In particular, this is the case for CTMRG, which has become a cornerstone of tensor network approaches in classical and quantum lattice models. This algorithm was first introduced by Nishino and Okunishi~\cite{Nishino1995,Nishino1996, Nishino1997} as an efficient contraction scheme to evaluate partition functions of infinite two-dimensional square lattice models. It comes as a combination of Baxter's corner transfer matrix~\cite{Baxter1968,Baxter1978,Tsang1979, Baxter1981} and White's density matrix renormalization group (DMRG) algorithm~\cite{white1992,white1993} . Although initially introduced in the context of 2D statistical physics as a complementary approach to other transfer matrix methods or to Monte Carlo, its use rapidly extended to 2D quantum physics. Indeed, in the last two decades, CTMRG has been used as a contraction algorithm for infinite projected entangled pair states (iPEPS) wavefunctions~\cite{Verstraete2004,Jordan2008, Orus2009,Cirac2021, Okunishi2022}. Pairing CTMRG with different update schemes for iPEPS such as simple~\cite{Jiang2008} and full update~\cite{fullupdate} and more recently automatic differentiation~\cite{Liao2019,Hasik2021} has notably supported new results in fermionic systems~\cite{Corboz2010, CorboztJ1, CorboztJ2,Zheng2017} and frustrated systems~\cite{Murg2009,Corboz2013,Hasik2019,Schmoll2023}.

\par Besides the now well-established power of tensor networks for quantum systems, a promise of tensor networks for classical spin systems has been to offer either a powerful alternative to Monte Carlo (see e.g. Refs.~\onlinecite{Vanderstraeten2018, Ueda2020,Nyckees2021,Schmoll2021,Nyckees2022HardSquare}) or a support to improve sampling~\cite{Viejira2021, FriasPerez2023}. However, in recent years, their application to frustrated two- and three-dimensional classical models has attracted some attention~\cite{Li2021,Wang2014,Zhu2019,Vanhecke2021,Liu2021,Colbois2022,Song2022,Song2023,Song2023_b,Akimento2023}, as it has been established that particular care has to be taken in the tensor network formulation to avoid numerical instabilities or convergence to wrong results. Importantly, these instabilities seem to get significantly reduced away from a macroscopically degenerate ground state~\cite{Li2021}, but they can create issues close to a critical ground state such as that of the triangular lattice Ising antiferromagnet~\cite{Vanhecke2021}. In the context of exact, differentiable contraction, these instabilities can be dealt with by working with the logarithm of the Boltzmann weight~\cite{Liu2021}. In the case of approximate contraction, where it is unclear how to use such a construction, a successful alternative approach has been to rely on ensuring that the ground-state local rule is satisfied at the level of the tensor, such that the low-temperature limit of the tensor network remains well-defined~\cite{Vanhecke2021,Colbois2022,Song2023_b}. This approach has been mostly validated using VUMPS; here, we find that it is also successful with various CTMRG algorithms.

\par Since CTMRG is defined on the square lattice, a usual first step is to map the tensor network from the original lattice onto the square lattice~\cite{Mangazeev2010,Corboz2012Spin,Corboz2012b,Jahromi2016,Jahromi2018,Bauer2012,Schmoll2023}. While it has been generally successful (of particular interest here, there is the case of the triangular lattice Ising ferromagnet,~\cite{Mangazeev2010}), this step may occasionally lead to problems such as poor convergence or non-physical breaking of symmetry~\cite{Lukin2023}; in contrast, it is well known that improved performance is achieved when the algorithm makes use of the underlying symmetry of the lattice~\cite{Vanderstraeten2022, Lukin2023}. In this spirit, proposing variants of existing algorithms formulated to fit the lattice symmetry of the problem could allow investigation of systems with better accuracy. 

In this paper, we design a CTMRG algorithm to contract infinite tensor networks defined on the honeycomb lattice, naturally giving rise to a $C_3$-symmetric contraction scheme. A simple construction enables us to
apply it to the low temperature phase diagram of the triangular antiferromagnet Ising model in a field. In Section ~\ref{sec:model} we describe this model and recall previous works on the topic. In Section~\ref{sec:CTM}, we recall the CTMRG algorithm on the square lattice and then introduce the CTMRG algorithm on the honeycomb lattice. In Section~\ref{sec:Results}, we first benchmark the algorithm on the classical Ising antiferromagnet on the triangular lattice, naturally formulated through a dual construction as a tensor network on the honeycomb lattice. We then revisit the effect of the magnetic field both in the constrained model and the finite-temperature case. We show that the location of the critical line at finite temperature can be evaluated with much higher precision than previously achievable. Finally, in Section~\ref{sec:Discussion} we discuss our results and provide an outlook.

\section{The model}
\label{sec:model}

\par The antiferromagnetic Ising model on the triangular lattice in a magnetic field is defined by the Hamiltonian: 
\begin{equation}
	\label{eq:H}
    \mathcal{H} = J \sum_{\langle i,j \rangle} \sigma_i \sigma_j - h \sum_{i} \sigma_i
\end{equation}
with $\sigma \in \{+1,-1 \}$ an Ising spin variable, $J>0$ the interaction parameter and $h$ the field parameter. 

\par
\begin{figure}[t!]
\includegraphics[width=0.8\columnwidth]{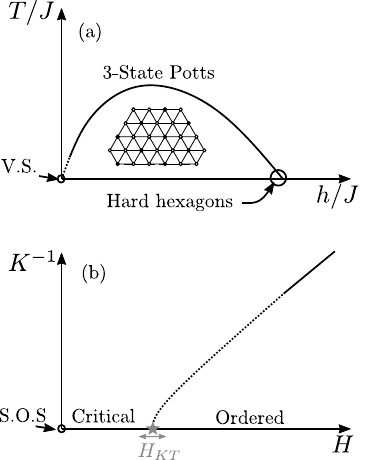}
\caption{(a) Sketch of the phase diagram in the $(h/J,T/J)$ units. The ordered phase melts through a three-state Potts transition. The slope of the critical line near the $m = 1$ magnetization plateau is determined by a mapping to the hard hexagon model. (b) Phase diagram of the model in reduced coordinates $(H,K^{-1})$ with $H = h/T$ and $K = J/T$. At $K^{-1} = 0$, the system maps onto an SOS model with a critical phase and an ordered phase separated by a KT transition.}
\label{fig:PDSketch}
\end{figure}

 We show a sketch of its phase diagram in Fig.~\ref{fig:PDSketch}. In the absence of a magnetic field, the model has a macroscopically degenerate ground-state~\cite{Wannier1950, Houtappel1950}, i.e. a finite residual entropy, and is characterized by each triangle satisfying a two-up one-down, two-down one-up (UUD/DDU) rule for the spins. This critical point has algebraically decaying spin-spin correlation~\cite{Stephenson1964} characterised by a critical exponent $\eta = 1/2$ and central charge $c = 1$, and is commonly referred to as the Villain-Stephenson (VS) point. It can be described using a Coulomb-gas construction~\cite{Nienhuis_1982,Nienhuis_1984, Nienhuis_1987} thanks to an exact mapping onto a triangular solid-on-solid (SOS) model~\cite{Blote_1982,Nienhuis_1984, Henley2010}. Upon introducing a positive field, the configurations with each triangle having UUD spins are favored, giving rise to long-range correlations. This $m=1/3$ magnetization plateau, with $\sqrt{3}\times \sqrt{3}$ symmetry breaking stabilizes until $h = 6J$. This point has a finite residual entropy $S = 0.333242...$~\cite{Baxter1980, Metcalf1978}. For $h > 6J$, the ground state is fully polarized. 

At zero field, for any nonzero temperature, the system has a finite correlation length which diverges exponentially fast as $T \rightarrow 0$~\cite{Stephenson1970,Jacobsen1997}. In contrast, for $0<h<6J$, the $\sqrt{3}\times \sqrt{3}$ ordered phase melts at finite temperature through a three-state Potts transition as expected from the ground-state symmetry~\cite{Alexander} and verified by phenomenological scaling and transfer matrix calculations~\cite{Kinzel1981,NohKim,TamashiroSalina,Blote91,Blote93,Qian}. This transition is characterized by critical exponents $\beta = 1/9, \nu = 5/6, \eta = 4/15, c = 4/5$, with $\beta$ describing the order parameter power law in the ordered phase and $\nu$ characterizing the divergence of the correlation length.

\par The shape of the Potts critical line in the limit $T \rightarrow 0$ has attracted quite some interest. In the vicinity of $h = 6J$, the shape of the line is predicted by the hard-hexagon model whose second-order phase transition is also in the three-state Potts universality class. Indeed, the partition function of the triangular Ising antiferromagnet in the limit $T\rightarrow0, h \rightarrow 6J$ with $(6J-h)/T =: \ln(\zeta)/2$ remaining finite maps to that of the hard hexagon model with fugacity $\zeta$~\cite{Racz,Qian}; thus the slope of the three-state Potts critical line in that limit is determined by the critical fugacity of the hard-hexagon model $\zeta_c = \frac{11+5\sqrt{5}}{2}$~\cite{Baxter1982} as
\begin{equation}
    T_c \rightarrow \frac{1}{\ln(\zeta_c)}(12J-2h)
\end{equation}

The limit $T\rightarrow0, h\rightarrow0$ has proven much more challenging. A first conjecture~\cite{Kinzel1981} was that the Potts line would approach the VS point with an infinite slope, but renormalization group investigation instead showed that in this limit, the slope of the transition line should remain finite~\cite{Nienhuis_1984}. The model is sometimes introduced considering the reduced Hamiltonian with parameters $H = h/T$ and $K = J/T$:
\begin{equation}
	\label{eq:HR}
    \mathcal{H}_R =  \frac{\mathcal{H}}{T} = K \sum_{\langle i,j \rangle} \sigma_i \sigma_j - H \sum_{i} \sigma_i
\end{equation}
where an additional critical phase appears at $K^{-1}=0$. Indeed, introducing a finite reduced field at the VS point acts as a perturbation which become relevant only when $\eta = 4/9$ where a Kosterlitz-Thouless (KT)\cite{Kosterlitz_Thouless,Jose1977,Nienhuis_1984} transition into an ordered phase must occur. The location of the KT transition is not predicted by the RG analysis, but transfer matrix studies have located it around $H_{KT} = 0.266\pm 0.01$~\cite{Blote91,Blote93,Qian,Otsuka06,Hwang2014}. The three-state Potts transition should then meet the KT point in the zero temperature limit. A study based on transfer matrix and renormalization group analysis~\cite{Qian} has given evidence in favor of that scenario and further suggested that the transition line in the $(K^{-1}, H)$ phase diagram approaches $H_{KT}$ with a square root singularity. However the smallest accessible critical field of the three-state Potts transition remains significantly larger than $H_{KT}$ and a definitive answer on the shape of the transition near $H_{KT}$ remains to be given.

For the rest of the paper, we will sometimes refer to $K^{-1}$ as the temperature but will always distinguish between the reduced field $H$ and the field $h$.

\section{Corner transfer matrix renormalization algorithm}
\label{sec:CTM}
In this section we give a brief overview of the square CTMRG and its implementation for the triangular Ising antiferromagnet. We then introduce a new version of the CTMRG algorithm that contracts honeycomb tensor networks and show how to express the partition function of the antiferromagnetic Ising model on the triangular lattice as the contraction of an infinite honeycomb tensor network.

\subsection{Algorithm for the square lattice}

\begin{figure}[t!]
\includegraphics[width=0.45\textwidth]{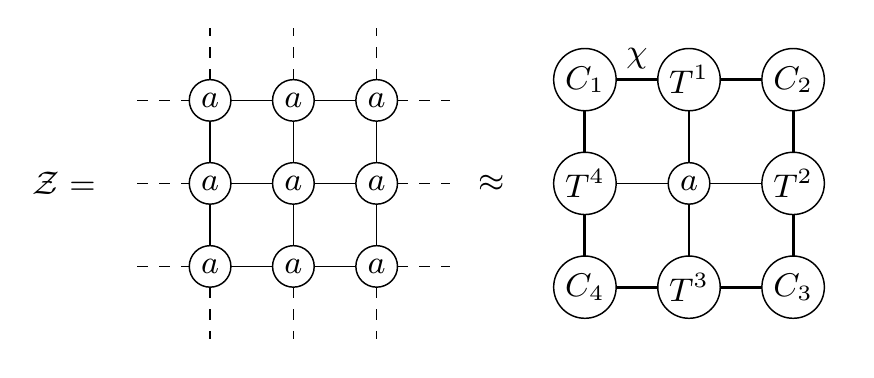}
\caption{The partition function written as the infinite contraction of a rank four tensor in a square lattice is approximated by a contraction of nine different tensors. Bold lines represent legs of bond dimension $\chi$ while thin lines represent bond dimension $d$.}
\label{fig:Zsquare}
\end{figure}

The square CTMRG algorithm approximates the contraction of infinite square tensor networks made of rank-four local tensor $a$ of dimension $d\times d\times d\times d$ with a contraction of eight different tensors referred to as the environment and the local tensor $a$ as shown in Fig. \ref{fig:Zsquare}. The environment is made of four corner tensors $C_i$ of dimension $\chi\times\chi$ and four edge tensors $T_i$ of dimension $\chi\times d\times \chi$. The approximation is controlled by the bond dimension $\chi$ and in the infinite bond dimension limit one recovers the exact result.

\begin{figure}
    \vspace{0.03\linewidth}
    \includegraphics[scale=0.5]{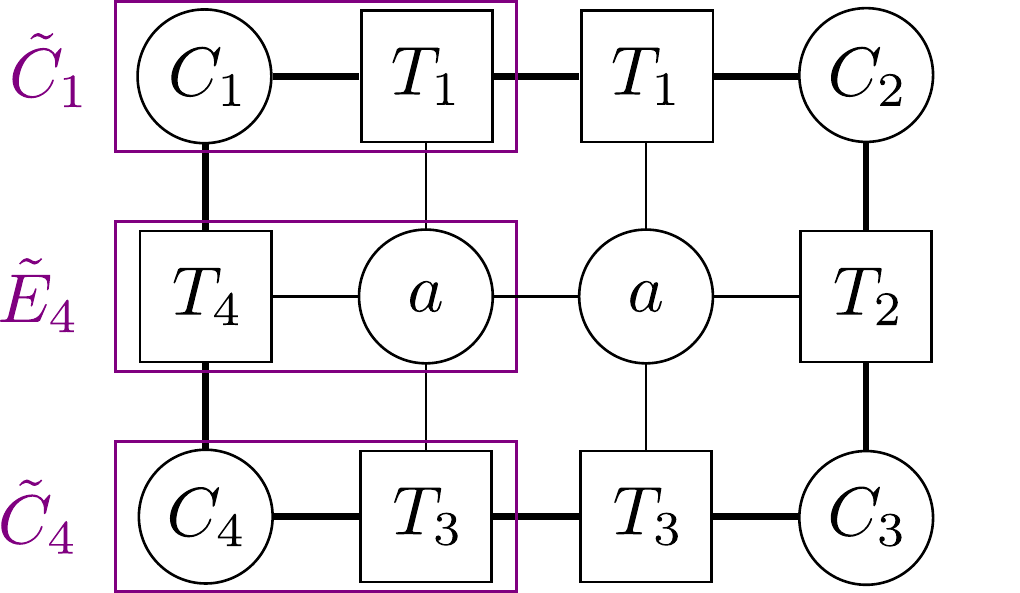}
    \caption{Illustration of the absorption step of square CTMRG in tensor network notation.}
    \label{fig:ctmrg_2}
\end{figure}

\begin{figure}
    \includegraphics[scale=0.5]{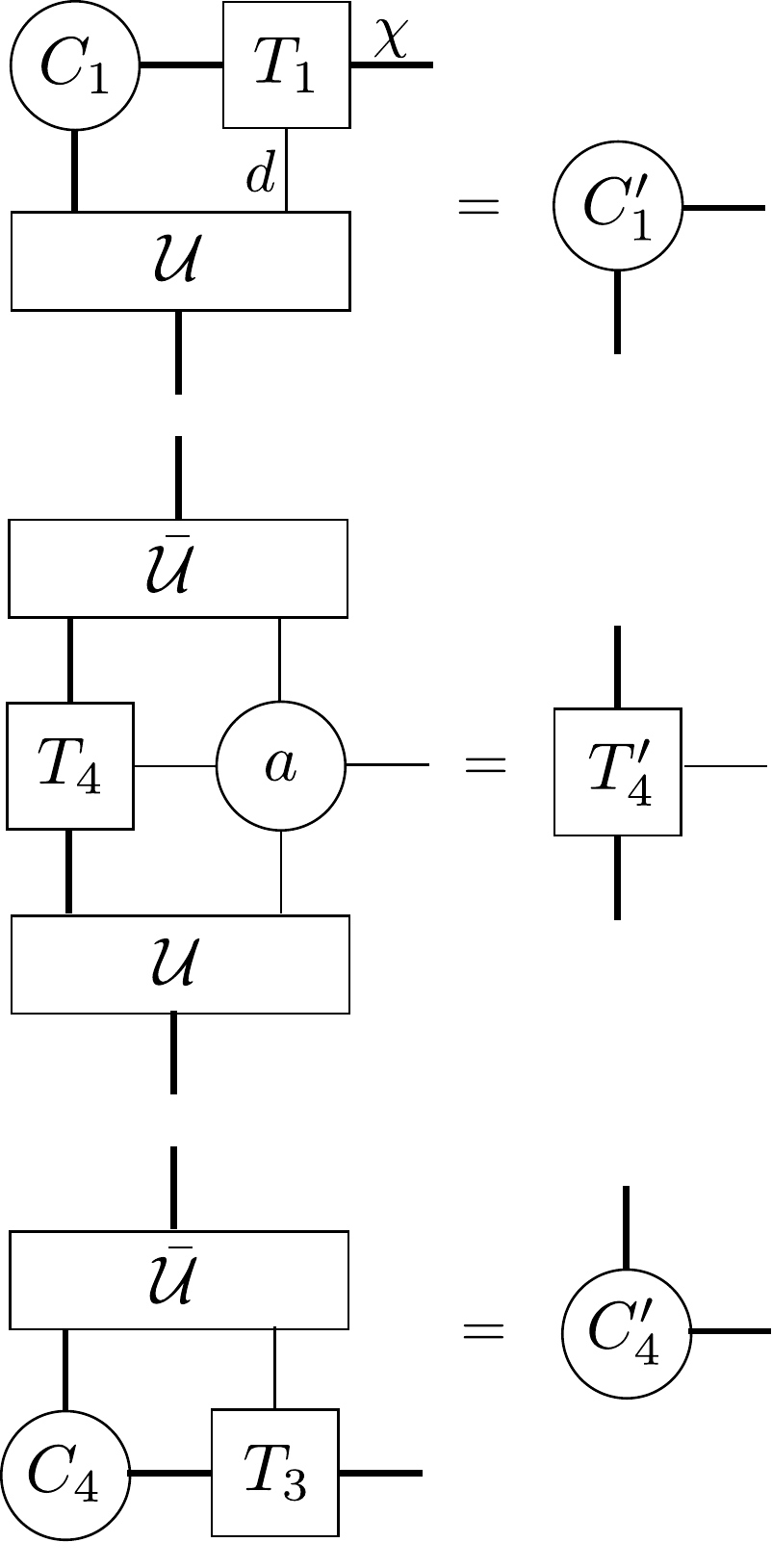}
    \caption{Illustration of the renormalization step of square CTMRG in tensor network notation.}
    \label{fig:ctmrg_3}
\end{figure}

In order to find a representation of the infinite lattice tensor network with a finite bond dimension, tensors are iteratively added into the network and the thermodynamic limit is obtained when some observables have converged. There are different ways of constructing the environment and in the present case we focus on the directional CTMRG algorithm \cite{Orus2009}, where columns are added to the network one at a time. It consists of two steps, which when repeated along the four lattice directions define a full CTMRG iteration:
\begin{itemize}
    \item Absorption: a new column is introduced into the network in one of the directions and contracted with the appropriate corner and edge tensors, increasing their dimension by a factor of $d$. An illustration of the absorption for a left move is included in Fig. \ref{fig:ctmrg_2}.
    \item Renormalization: if one of the dimensions of a tensor exceeds the cutoff dimension, the dimension of this index is truncated after application of an isometry $\mathcal{U}$. This step is illustrated in Fig. \ref{fig:ctmrg_3} for a left move. 
\end{itemize}
There are several possible choices for the definition of the isometry $\mathcal{U}$, which can affect the convergence of the algorithm. In the present case we used the isometry first introduced by Corboz in \cite{IsoCorboz}. \footnote{The isometries first introduced by Nishino and Okunishi in Ref. \cite{Nishino1995} or Orus in Ref. \cite{Orus2009} performed poorly at small reduced field and close to the critical regime.}

One of the advantages of the CTMRG formalism is that it gives access to the entanglement entropy, whose divergence at second order phase transitions can be used to study the associated universality classes~\cite{Chatelain2020,Ueda2020,Gendiar6statePotts2} and is defined as :
\begin{align}
        S & = -\text{Tr}(\rho \log(\rho))
\end{align}
by setting the density matrix $\rho$ as:
\begin{align}
        \rho & = C_1C_2C_3C_4/\text{Tr}(C_1C_2C_3C_4).
\end{align}

\subsubsection*{Mapping on the triangular lattice}

\begin{figure}[t!]
\includegraphics[width=0.45\textwidth]{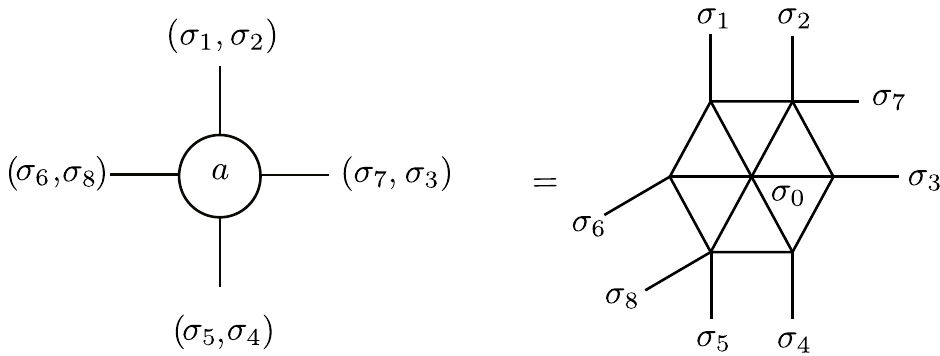}
\caption{Local tensor made of seven spins given by Eqs.~\eqref{eq:square_tensor} and~\eqref{eq:local_H}.}
\label{fig:square_ctm_tensor}
\end{figure}

For a given statistical mechanical system the choice of a tensor network representation of its partition function is not unique. In the case of frustrated systems such as the triangular Ising antiferromagnet the choice of the tensor greatly affects the convergence. It was observed \cite{Vanhecke2021,Song2022} that it is best to use a tensor network construction obtained by factorizing the Boltzmann weight of a global spin configuration into a product over a set of tiles whose reunion covers the whole lattice. In this spirit, we define a tensor for the triangular Ising antiferromagnet that accounts for the Boltzmann weight of a hexagon of seven spins given by :
\begin{align}
    a_{(\sigma_1,\sigma_2),(\sigma_7,\sigma_3),(\sigma_5,\sigma_4),(\sigma_6,\sigma_8)}=\sum_{\sigma_0=\pm1}e^{-\beta H(\sigma_0,\dots,\sigma_6)}\delta_{\sigma_2,\sigma_7}\delta_{\sigma_5,\sigma_8}
    \label{eq:square_tensor}
\end{align}
with 
\begin{multline}
 \label{eq:local_H}
    H(\sigma_0,\dots,\sigma_6) =  J\left(\sigma_0\sum_{i = 1}^6 \sigma_i +  \sum_{i = 1}^{5} \sigma_i \sigma_{i+1} + \sigma_6 \sigma_{1}\right)\\
     - h \sigma_0  - \frac{h}{3}\sum_{i = 1}^{6} \sigma_i
\end{multline}
We give the schematic picture of the tensor in Fig. \ref{fig:square_ctm_tensor}.  

\subsection{Algorithm for the honeycomb lattice}

We now introduce a variant of CTMRG that contracts infinite tensor networks defined on the (bipartite) honeycomb lattice with local tensors $a$ and $b$ of dimension $d\times d\times d$. It is worth noting that there are other CTMRG algorithms on lattices other than square~\cite{Daniska2015,Daniska2016}. The most relevant in our case has originally been introduced by Gendiar \textit{et al.}~\cite{ Gendiar2012} while studying Ising models on triangular-tiled hyperbolic lattices formulated as interactions-round-a-face (IRF) tensor networks. This was recently adapted to the context of 2D-quantum physics model using automatic differentiation~\cite{Lukin2023} with iPEPS wave-functions defined on a honeycomb lattice with a single local tensor $a$. Although they apply to similar lattices, we see a key difference between this algorithm and the one presented here. In Ref.~\onlinecite{Lukin2023}, the corner matrix represents a sixth of the partition function while in ours it corresponds to a third. This in turn leads to inserting the isometries in different places and to a more natural generalization of our algorithm when the bipartite lattice is made of two different local tensors. We postpone this discussion to the end of the present section. 

\begin{figure}[t!]
\includegraphics[width=0.45\textwidth]{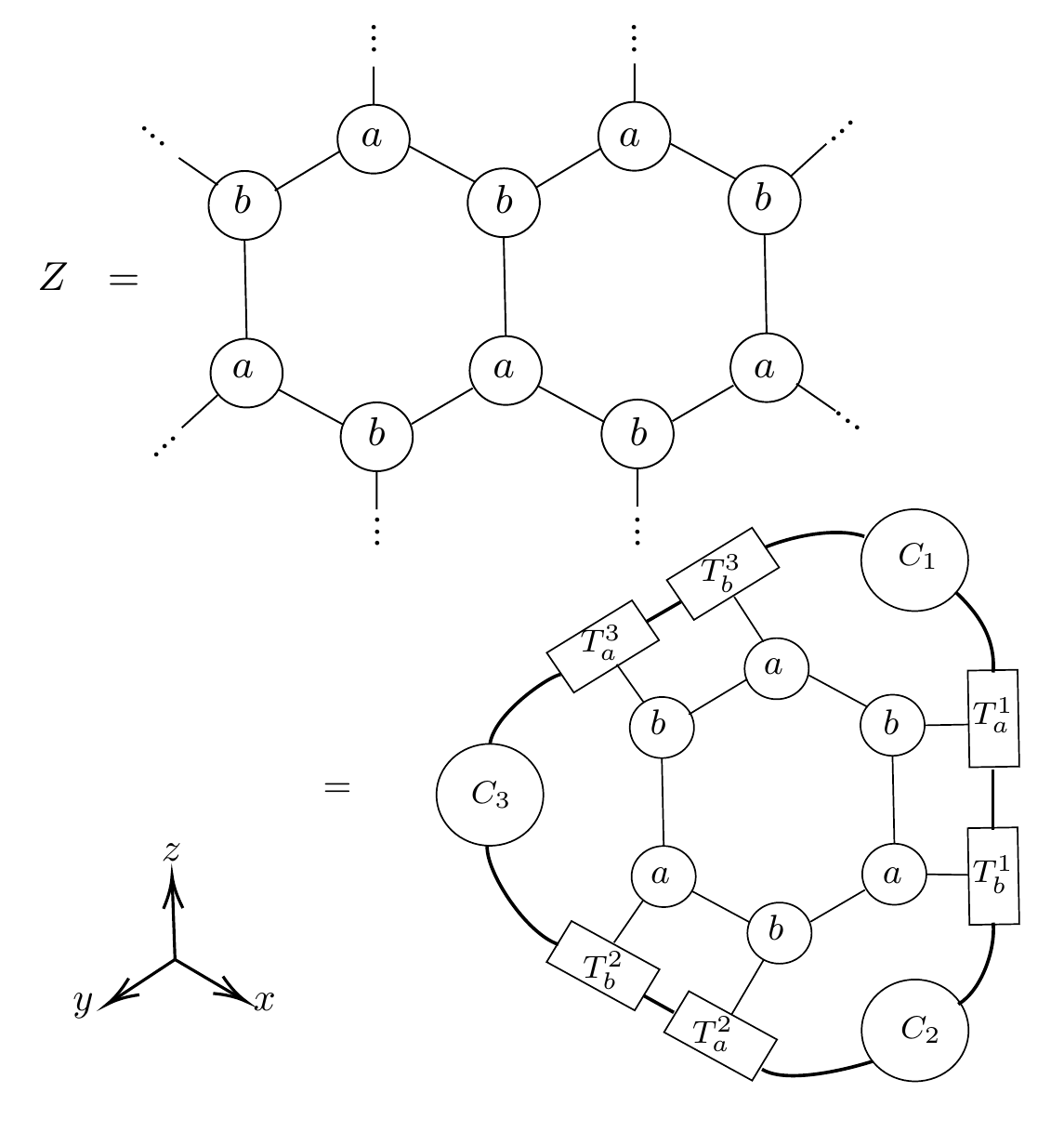}
\caption{Partition function written as a two dimensional honeycomb tensor network, and then contracted to an environment made of nine tensors, and a unit cell made of six on-site tensors. Bold lines represents bond dimension of $\chi$ while thinner lines indicate a smaller bond dimension of $d$.}
\label{fig:honecombZ}
\end{figure}

\begin{figure}[t!]
\includegraphics[width=0.40\textwidth]{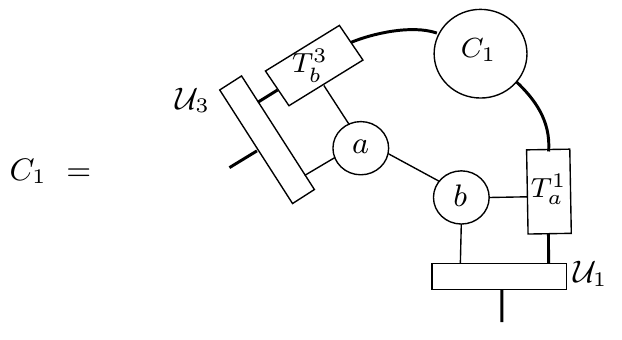}
\caption{One full iteration for the corner tensor $C_1$, Eqs.~\eqref{eq:Cu} and~\eqref{eq:Ct}. $C_2$ and $C_3$ are updated similarly}
\label{fig:evolc}
\end{figure}

\begin{figure}[t!]
\includegraphics[width=0.40\textwidth]{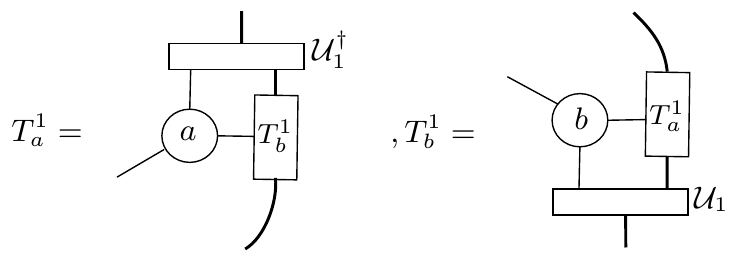}
\caption{One full iteration for the edge tensors $T_a^1$ and $T_b^1$, Eqs.~(\ref{eq:Tau},~\ref{eq:Tbu},~\ref{eq:Tbt},~\ref{eq:Tat}). The other edge tensors are updated similarly.}
\label{fig:evoltt}
\end{figure}

\begin{figure}[t!]
\includegraphics[width=0.40\textwidth]{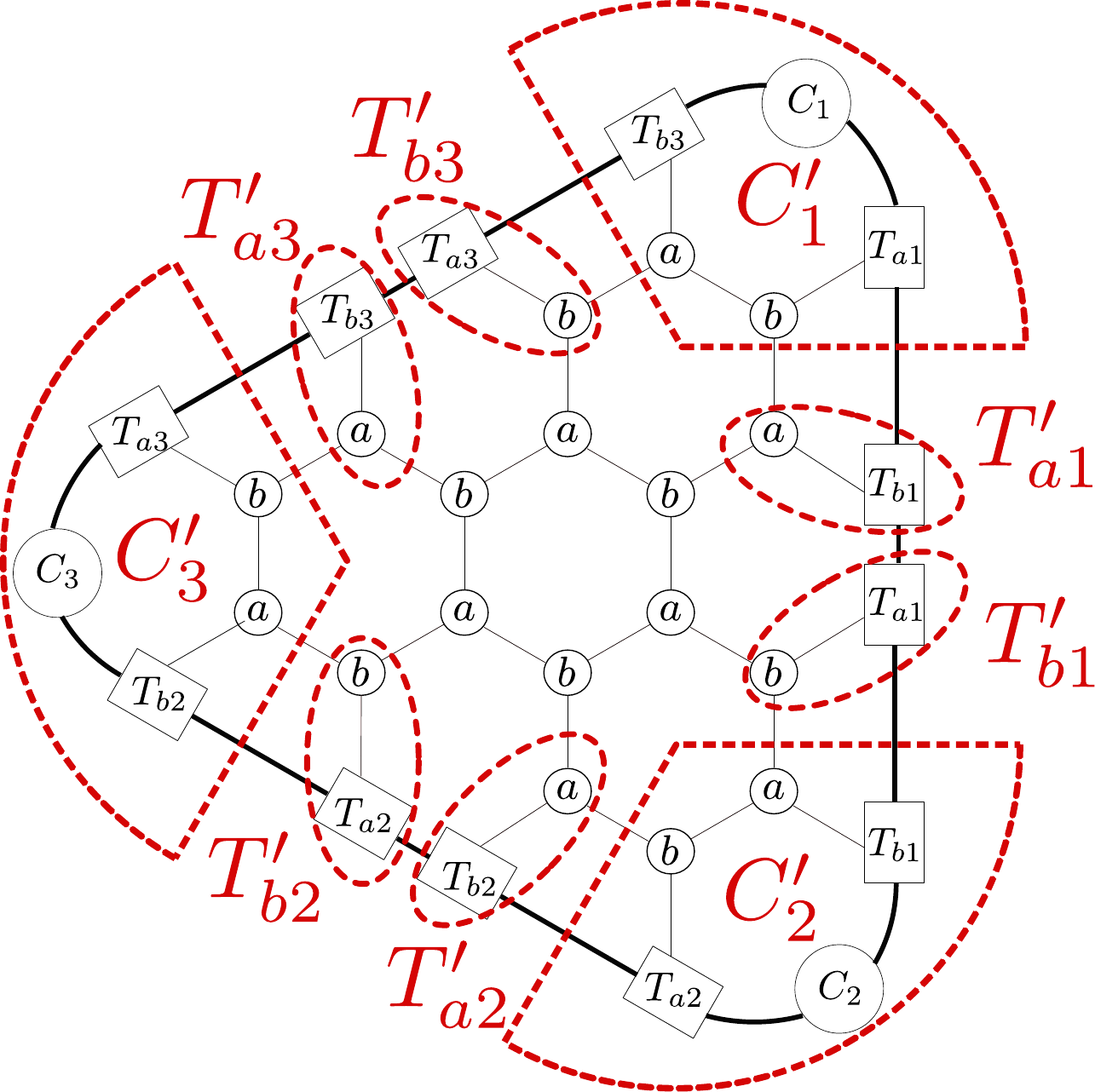}
\caption{Schematic way of representing how the environment is updated. Primed tensors represent the tensors after one iteration.}
\label{fig:expl_ctmrg}
\end{figure}

The algorithm presented here approximates the infinite tensor network with nine tensors which represent the environment $E = \{T_\alpha^i, C_i \mid i \in \{1,2,3\}, \alpha\in \{a,b\}\}$ as shown in Fig.~\ref{fig:honecombZ} where the corner tensors $C_i$ are of dimension $\chi \times\chi$, and the edge tensors $T_{\alpha}^i$ are of dimension $\chi \times d \times \chi$. Just as in the square lattice CTMRG case~\cite{Nishino1996}, the approximation is controlled by the bond dimension $\chi$ and in the infinite $\chi$ limit, one recovers the exact result. The algorithm iterates two alternating steps as well, the update and the renormalization, until the convergence of some observable of interest\footnote{See e.g. Refs.~\onlinecite{Okunishi2022,Nishino1996Numerical} for a discussion of finite-size vs finite-bond dimension in the square lattice CTMRG.}. 
For the CTMRG on the honeycomb lattice, we propose the following update (the full iteration is shown in Figs.~\ref{fig:evolc} and~\ref{fig:evoltt}):

\begin{align}
     C_{i}' & = abC_{i}T_{a}^{i} T_{b}^{f(i-1)} \label{eq:Cu}\\
     T_{b}^{i'} & = T_{a}^{i} b \label{eq:Tau}\\
     T_{a}^{i'}& = a T_{b}^{i}  \label{eq:Tbu}
\end{align}
where we have introduced $f(i) = \text{mod}(i,3)$ to make the notation lighter. We then have to project the tensors in a relevant subspace in order to keep their dimensions under control. 
This is called the renormalization step and goes as:

\begin{align}
    C_i & =  \mathcal{U}_i C_i'  \mathcal{U}_{f(i-1)}^\dagger \label{eq:Ct}\\
    T_{b}^{i} & = T_{b}^{i'} \mathcal{U}_i \label{eq:Tbt} \\
    T_a^i & = \mathcal{U}_i^\dagger  T_a^{i'} \label{eq:Tat}
\end{align}
where the isometry $\mathcal{U}_i$ is computed by truncating the singular value decomposition of $C_{f(i+1)}C_{f(i+2)}C_i$. This isometry would be the equivalent in the square CTMRG of the isometry originally used by Nishino and Okunishi~\cite{Nishino1996}. We illustrate picturally in Fig. \ref{fig:expl_ctmrg} how the iterations correspond to contracting the partition function on the whole honeycomb lattice.

Similarly to the square CTMRG we can compute the entanglement entropy by setting the density matrix $\rho$ as:
\begin{align}
        \rho & = C_1C_2C_3/\text{Tr}(C_1C_2C_3).
\end{align}

A particularly useful case is when the local tensors $a$ and $b$ are invariant by rotation. Then, one can reduce the environment to only three different tensors, $E = \{C,T_a,T_b\}$ and the density matrix becomes:
\begin{align}
        \rho & = C^3/\text{Tr}(C^3)
\end{align}
We will refer to the algorithm making use of that symmetry as the $C_3$ symmetric (honeycomb) CTMRG. 

In the case of local tensors which are symmetric under rotation and reflection (without any particular relation between $a$ and $b$), the $C_3$ symmetric CTMRG algorithm gives direct access to the transfer matrix whose spectrum can be used to compute the correlation lengths and wave-vector in their respective directions. Indeed, by denoting the normalized eigenvalues of the transfer matrix as:
\begin{align}
	\lambda_i = e^{-\epsilon_i + i\phi_i}
\end{align}
one can show that the correlation length and wave-vector are given by:
\begin{align}
	\xi_{\chi} = \epsilon_2^{-1} & \qquad q_{\chi} = \phi_2.
\end{align}

\begin{figure}[t!]
\includegraphics[width=0.30\textwidth]{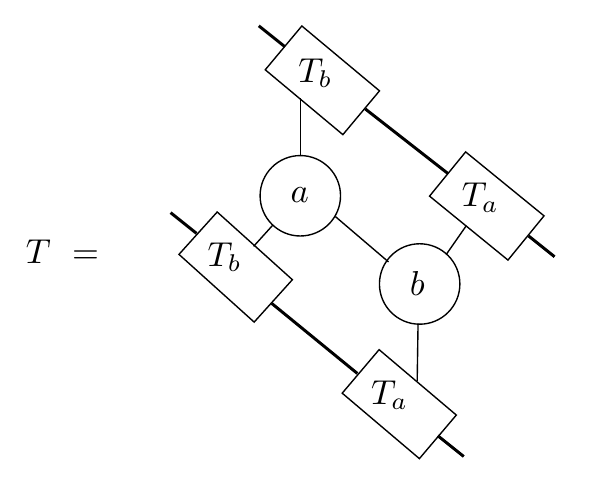}
\caption{Transfer matrix in the $x$ direction in the case of fully symmetric tensors $a$ and $b$. }
\label{fig:tm1}
\end{figure}

It was suggested by Rams~\textit{et al.}~\cite{PRXRams} to extrapolate the finite bond dimension correlation length with respect to higher gaps $\delta$ in the transfer matrices as:
\begin{align}
 \xi_{\chi}^{-1} = \xi^{-1}_{\infty} + a \delta 
\end{align}
Indeed, when the bond dimension increases, the spectrum above the first gap needs to converge to a continuum and $\delta$ goes to zero. We will be using in the next section $\delta = \epsilon_4 - \epsilon_2$. In Fig.~\ref{fig:tm1}, we show the transfer matrix in the $x$ direction with the convention of direction shown in Fig.~\ref{fig:honecombZ}.

Furthermore, if the infinite corner tensor is hermitian, one can keep $C$ diagonal and use an eigenvalue decomposition for better numerical stability. In particular, this is the case if the tensor network has a $C_{3v}$ symmetry.

We now briefly discuss the practical differences between the algorithm introduced in Ref.~\onlinecite{Lukin2023} and the $C_3$-symmetric CTMRG introduced here. As already mentioned, the main difference between the two algorithms is the definition of the corner transfer matrix which in turn leads to a different truncation scheme. In the present case we consider the CTM to represent a third of the partition function while in Refs.~\onlinecite{Lukin2023} it represents a sixth. Using Gendiar's algorithm on a bipartite lattice, one would then need to define two different corners $C_a$ and $C_b$ which in turn at the renormalization step would naively require two different isometries $U_1$ and $U_2$ to be projected onto their relevant subspace such that $C'_a = U_1C_aU_2$ and $C'_b = U_2 C_b U_1$. In contrast using only one corner matrix $C$ which effectively represents $C = C_aT_a T_bC_b$ only requires one isometry and $C' = UCU$.

\subsubsection*{Mapping from the triangular to the honeycomb lattice}

\begin{figure}[t!]
\includegraphics[width=0.35\textwidth]{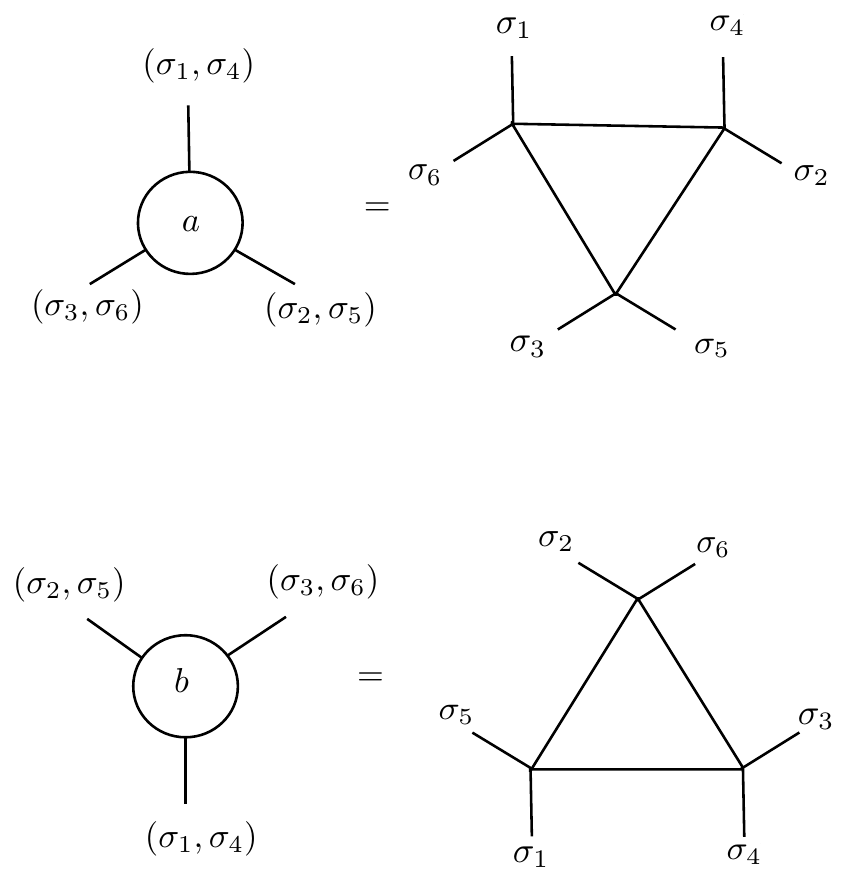}
\caption{Definition of the local tensors $a$ and $b$. By grouping the indices pairwise we end up with tensors $a$ and $b$ of dimension $4\times 4\times4$. }
\label{fig:localtens}
\end{figure}

\begin{figure}[t!]
\includegraphics[width=0.5\textwidth]{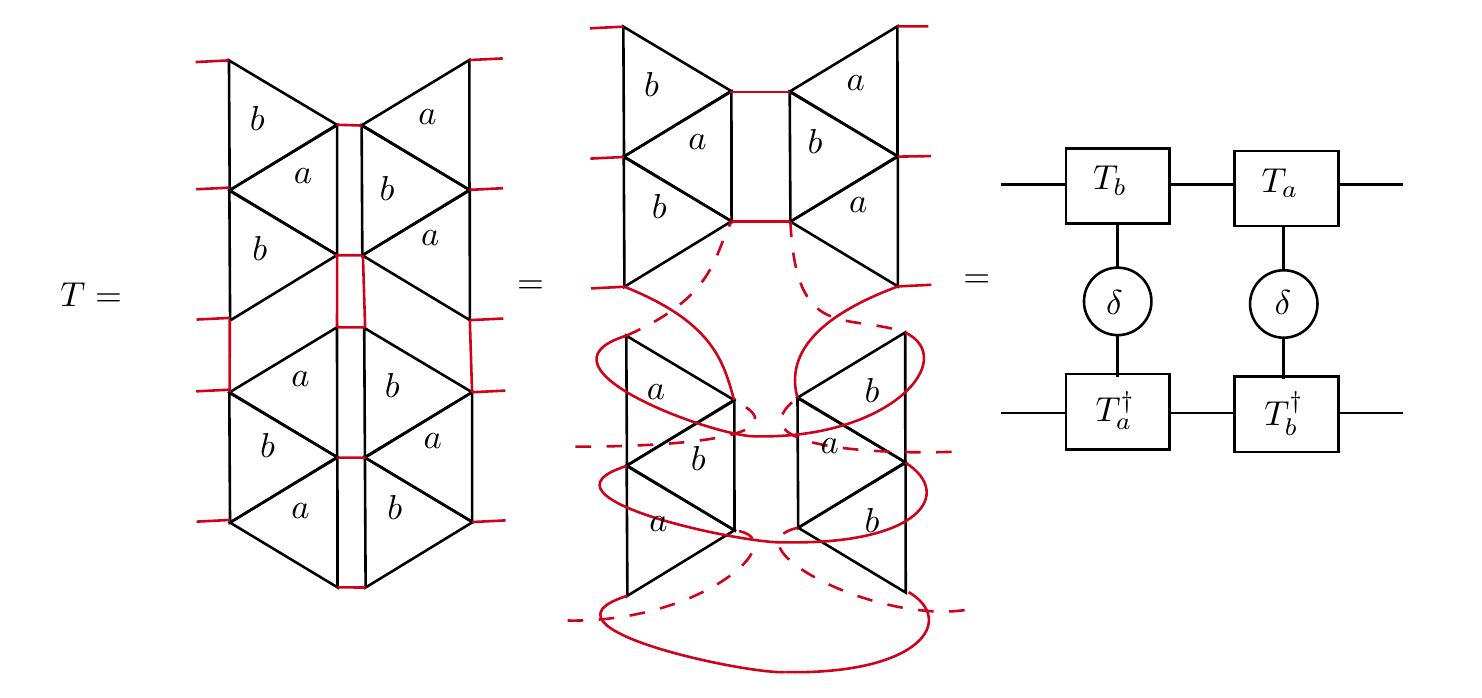}
\caption{Transfer matrix in the $x$ direction for the Ising model for a $6\times \infty$ system on the triangular lattice. The same reasoning applies to an infinite system in both directions and give the same results. The red lines indicate lines that must be contracted. The dashed lines represent the perspective in 3D and indicate that the legs pass below the full lines.}
\label{fig:tm4}
\end{figure}

To apply the $C_3$-symmetric CTMRG algorithm to the triangular lattice Ising antiferromagnet we map the triangular lattice onto the honeycomb lattice by defining the local tensors $a$ and $b$ on the dual as:
\begin{align}
		\label{eq:ab}
       a_{(\sigma_{1},\sigma_{4})(\sigma_{2}, \sigma_{5})(\sigma_{3}, \sigma_{6})} & =  \delta_{\sigma_1 \sigma_6} \delta_{\sigma_4 \sigma_2} \delta_{\sigma_3 \sigma_5}e^{\frac{H}{6}(\sigma_{1}+\sigma_{2}+\sigma_{3}) }  \nonumber \\
      & \times  e^{-\frac{K}{2}}  e^{-\frac{K}{2}(\sigma_{1}\sigma_{2}  +\sigma_{2}\sigma_{3}+\sigma_{3}\sigma_{1})}\\
           b_{(\sigma_{1} ,\sigma_{4})(\sigma_{2}, \sigma_{5})(\sigma_{3}, \sigma_{6})}& =   \delta_{\sigma_1 \sigma_5} \delta_{\sigma_2 \sigma_6} \delta_{\sigma_4 \sigma_3}e^{\frac{H}{6}(\sigma_{1}+\sigma_{2}+\sigma_{3}) }  \nonumber \\
      & \times  e^{-\frac{K}{2}}  e^{-\frac{K}{2}(\sigma_{1}\sigma_{2}  +\sigma_{2}\sigma_{3}+\sigma_{3}\sigma_{1})} \nonumber 
\end{align}
where $\delta$ denotes the Kronecker delta with $K$ and $H$ the reduced coupling constants. We give a diagrammatic expression of the tensors $a$ and $b$ in Fig.~\ref{fig:localtens}. A more detailed explanation on how to express a classical partition function as the contraction of a two-dimensional network and how to map the problem on a specific lattice geometry can be found in ~\cite{Vanhecke2021}. Defined as such, the local tensors $a$ and $b$ are rotation invariant, and we can use the $C_3$ CTMRG algorithm. We note that the tensors $a$ and $b$ are related by $a_{i,j,k} = b_{k,j,i}$. Although the tensors $a$ and $b$ are not fully symmetric and the previous construction of the transfer matrix (Fig. \ref{fig:tm1}) cannot be applied, we can still construct a transfer matrix but in a different way as shown in Fig. \ref{fig:tm4} with $\delta$ a $4\times 4$ matrix representing the permutation of two legs given by:
\begin{align}
    \delta =&  
    \begin{pmatrix}
        1 & 0 & 0 & 0 \\
        0 & 0 & 1 & 0 \\
        0 & 1 & 0 & 0 \\
        0 & 0 & 0 & 1
    \end{pmatrix}
\end{align}

This construction applies to the case of the triangular Ising model and does not generalize to arbitrary rotational invariant tensors $a$ and $b$.

\begin{figure}[t!]
\includegraphics[width=0.35\textwidth]{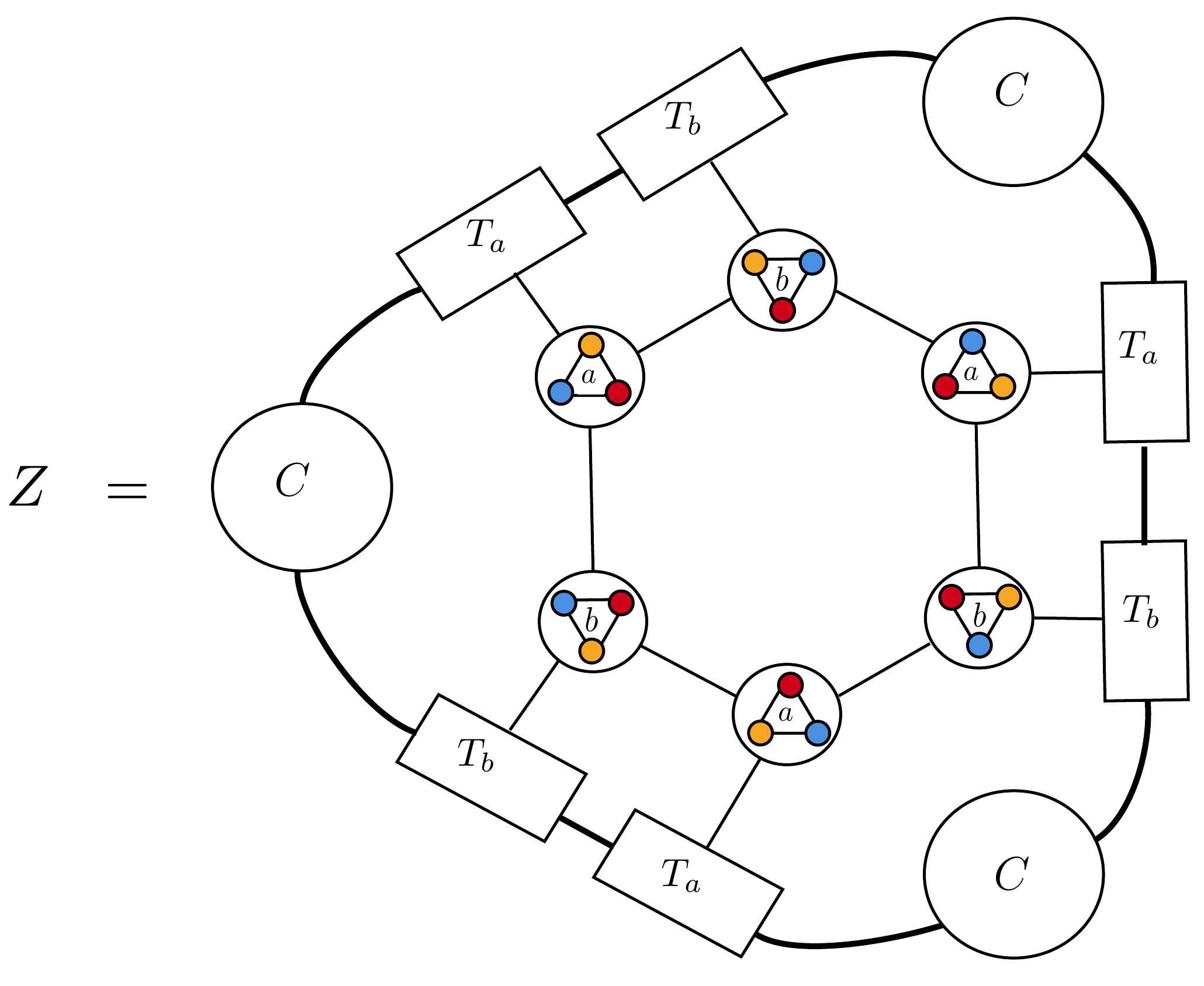}
\caption{Illustration of the implication of the $C_3$ symmetry on the expectation value of local observables. The tensors $a$ and $b$ are given by Eq.~\eqref{eq:ab}, and each tensor describes three sites. The $C_3$ symmetry in the environment imposes that different sites on the same sub-lattice have the same local magnetization. The figure highlights the sub-lattices on the tensors $a$ and $b$. One site belongs to the sub-lattice A (\protect\tikz \protect\draw[rouge,fill = rouge] (0,0) circle (.5ex);). Three sites belong to the sub-lattice B (\protect\tikz \protect\draw[bleu,fill = bleu] (0,0) circle (.5ex);) and three more sites belong to the sub-lattice C (\protect\tikz \protect\draw[orange,fill = orange] (0,0) circle (.5ex);).}
\label{fig:z3inv}
\end{figure}

In the ordered phase, the system undergoes a translational symmetry breaking and the triangular lattice is divided into three sub-lattices $A,B$ and $C$ with different magnetizations (Fig.~\ref{fig:PDSketch}).
We note that the honeycomb CTMRG naturally allows for this freedom while imposing the translational symmetry associated with the long-range correlations in the ground state. This property can be deduced by looking at Fig.~\ref{fig:z3inv}, where we have highlighted the different sub-lattices in the local tensors $a$ and $b$. Indeed, any observable measured on a site of the sub-lattice $A,B$ or $C$ (represented by colored dots) has to be the same if measured on a different site of the same sub-lattice due to the network being invariant by $C_3$. This is only valid when the corner and edge tensors are identical in all three possible directions. As we will see, enforcing this symmetry will lead to a better convergence and will allow us to access lower temperatures than previously available.

\section{Results}
\label{sec:Results}
We now present the results obtained by the $C_3$-symmetric CTMRG. We benchmark the algorithm on the antiferromagnetic triangular Ising model at zero and finite temperature. We then turn to the constrained model, where we investigate the location of the KT transition. Finally, we discuss and compare results obtained from the square and honeycomb CTMRG. We define the order parameter as:
\begin{align}
    \psi = \frac{1}{3} | \psi_A + e^{2\pi i/3}\psi_B + e^{-2\pi i/3}\psi_C |
\end{align}
with $\psi_i$ the magnetization on the sub-lattice $i$. In the ordered phase, due to the $\sqrt{3}\times \sqrt{3}$ unit cell, the honeycomb CTMRG algorithm converges to three different environments $E_A,E_B$ and $E_C$. Observables such as the magnetization in the middle of the plaquette or the entanglement entropy thus converge modulo three. Similarly, the transfer matrices obtained from $E_i$'s do not have the same spectrum. One cannot then consider this as the proper transfer matrix and would need to take into account all environments in order to construct the true transfer matrix. To do so, we would need to generalize the algorithm to a CTMRG with a general, multi-site unit cell~\cite{Orus2009}. We thus only have access to the correlation length in the disordered phase. Yet, as we will see, a two-site unit cell CTMRG is powerful enough for our purposes.

\subsection{Benchmark at zero field}

\begin{figure}[t!]
\includegraphics[width=0.45\textwidth]{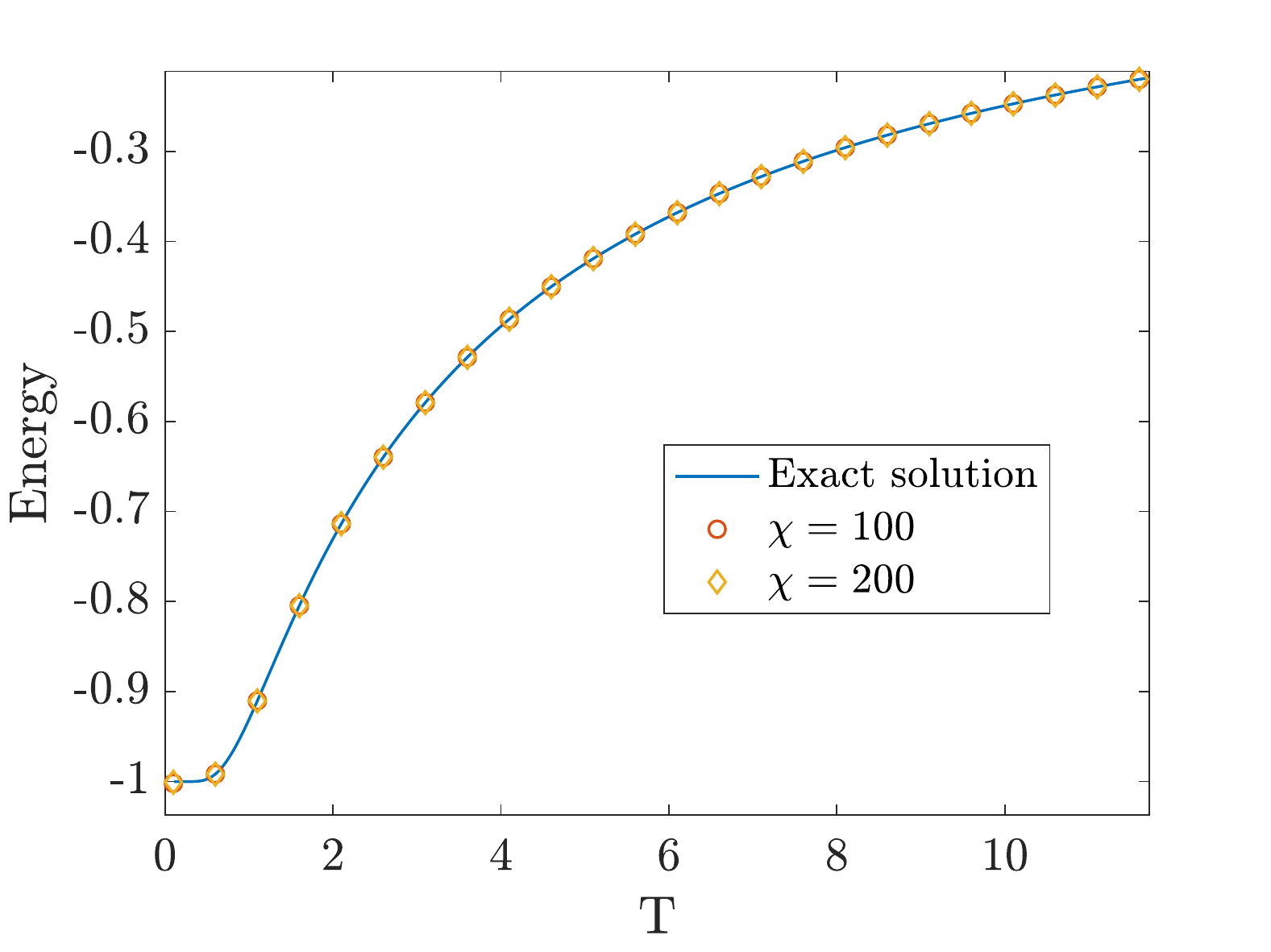}
\caption{Simulations were done at $h = 0$ for two different bond dimensions. The error is of the order $10^{-13}$ for temperatures larger than 0.6 where the correlation length is small. For the smallest computed temperature ($T = 0.1$), the error is of the order $10^{-10}$.}
\label{fig:enerh00}
\end{figure}

\begin{figure}[t!]
\includegraphics[width=0.45\textwidth]{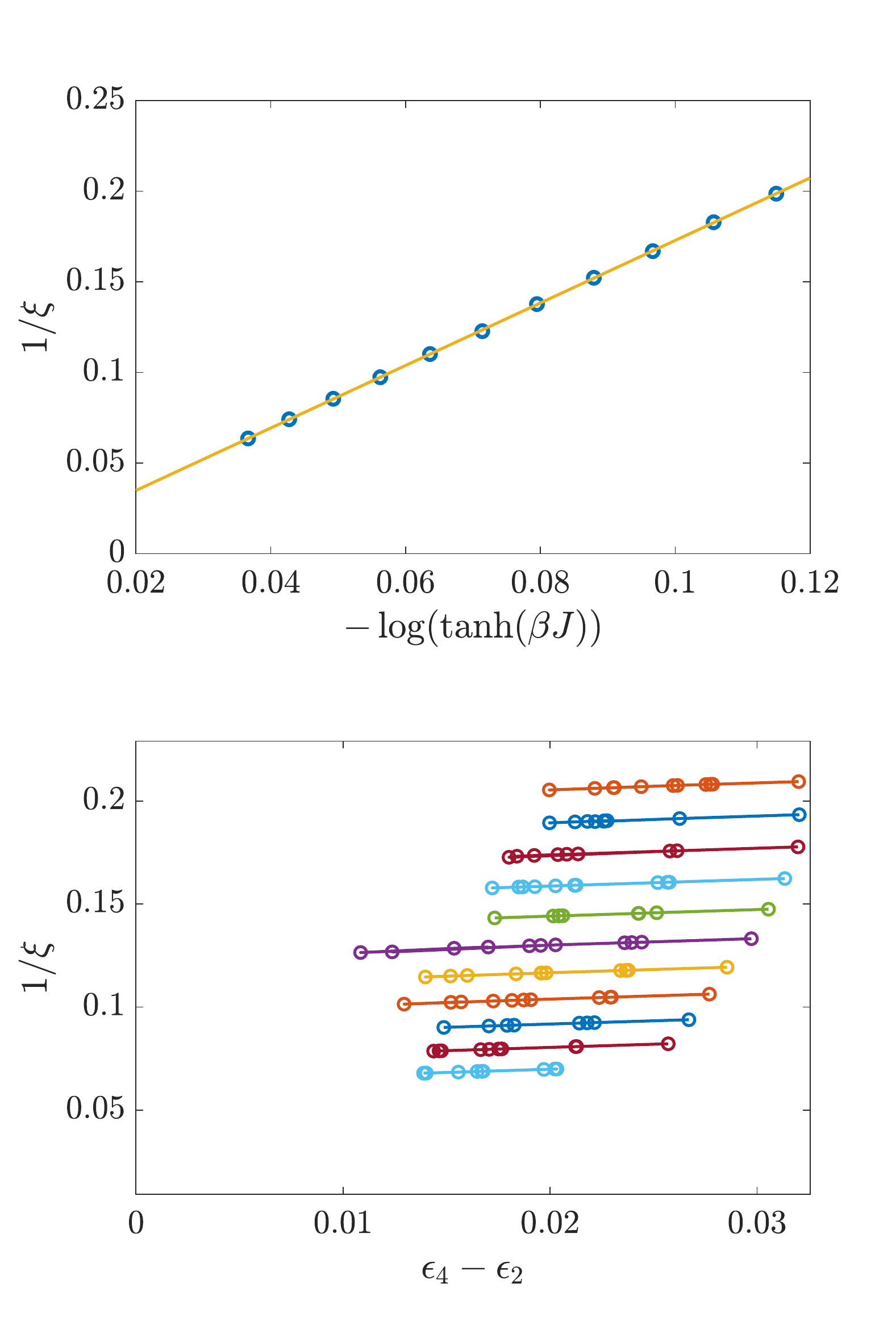}
\caption{Simulation done at $h = 0$. Top panel : the correlation length with respect to the temperature. The straight line represents a linear fit of the inverse correlation length. Bottom panel : the extrapolation used for the correlation length with respect to the gap $\delta = \epsilon_4 - \epsilon_2$ in the transfer matrix. We have used bond dimensions $\chi \in[100,200]$.}
\label{fig:corr00}
\end{figure}

As mentioned in the introduction, at $h= 0$ we recover the solvable triangular lattice Ising antiferromagnet whose energy is exactly known~\cite{Wannier1950,Wannier1973,Houtappel1950}. Its ground state has critical correlations, and upon decreasing the temperature, the correlation length is known to diverge exponentially fast~\cite{Stephenson1970,Jacobsen1997} as :
\begin{align}
\frac{1}{\xi} = -\log(\tanh(\beta J))
\end{align}
and this critical point belongs to the Villain-Stephenson universality class~\cite{Stephenson1970,Villain1977,Tanaka1992}. We compare our results with the exact values in Fig.~\ref{fig:enerh00} and \ref{fig:corr00} where the energy and correlation length are in good agreement with the associated theoretical predictions. 

\begin{figure}[t!]
\includegraphics[width=0.45\textwidth]{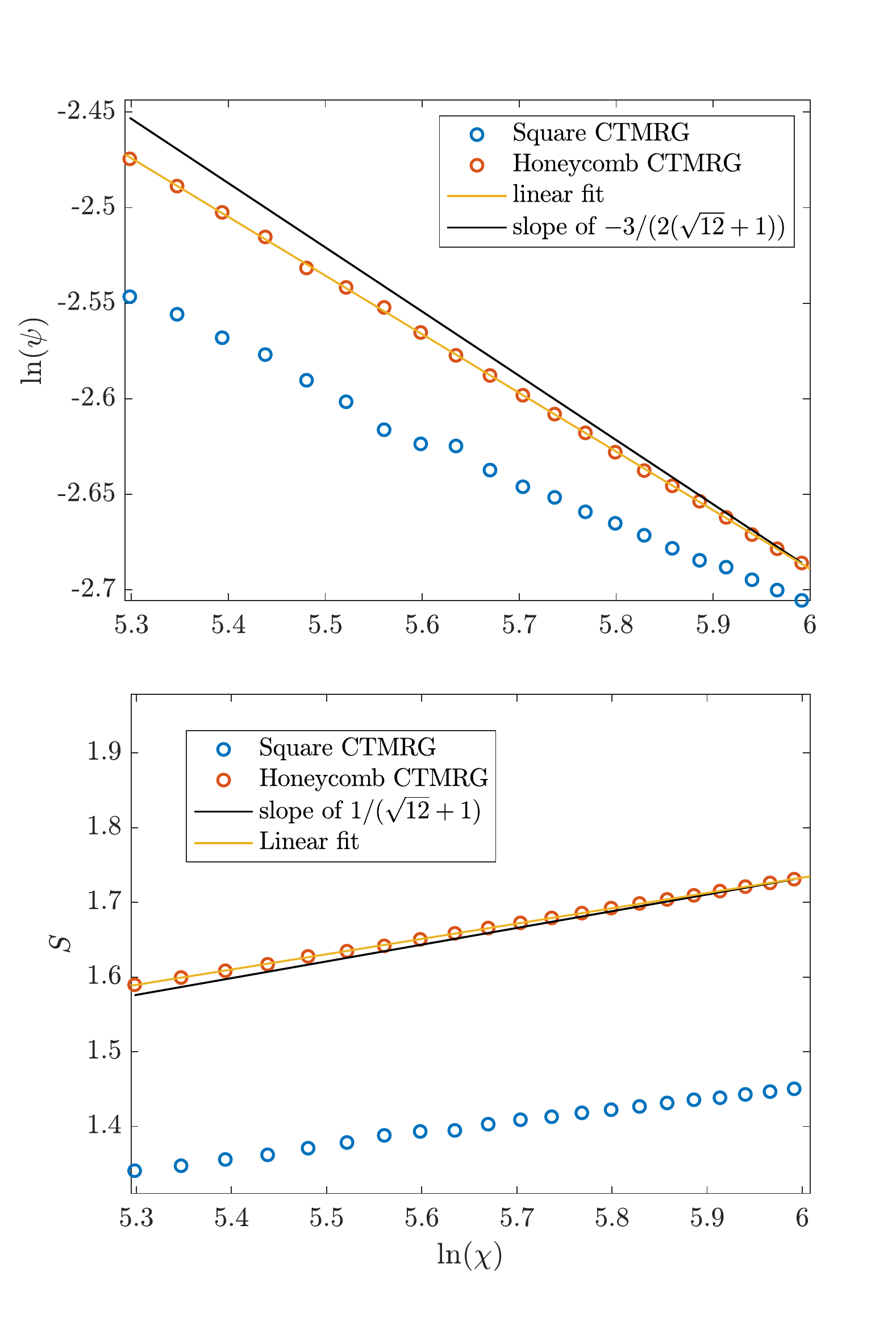}
\caption{Simulations done at $h = 0$ and $T = 0$ with the square and honeycomb CTMRG algorithms using $\chi\in[200,400]$. Top panel : the logarithm of the order parameter versus the logarithm of the bond dimension. Bottom panel : entanglement entropy versus the logarithm of the bond dimension. Results obtained from the honeycomb CTMRG are more accurate than the one obtained from the square CTMRG. The black lines indicate the slope predicted by Eq. \ref{eq:kappa} using $c = 1$ while the yellow lines are the linear fits from which we extract the critical exponents.}
\label{fig:comparison}
\end{figure}

At zero field, the system is critical with infinite correlation length. However, when introducing a finite bond dimension $\chi$, the correlation length will become finite as well and can only be increased by considering a larger bond dimension. It has been argued that it follows a power law $\xi \sim \chi^{\kappa}$ with universal exponent $\kappa$ given by~\cite{Pollmann2009,TagliaCozzo2008}:
 \begin{align}
     \kappa = \frac{6}{c(\sqrt{12/c}+1)}.
  \label{eq:kappa}
 \end{align}
Using the correlation length scaling, one can derive the following finite bond dimension dependency of the order parameter and entanglement entropy~\cite{Pollmann2009, TagliaCozzo2008} :
\begin{align}
\label{eq:scale1}  
    & \psi =   \chi^{-\frac{\kappa\eta}{2}} \\
     & S = \frac{c\kappa}{6} \log(\chi). 
\label{eq:scale2}  
\end{align}
Assuming the central charge, we can thus define effective exponents $\kappa$ and $\eta$ by fitting the order parameter and entanglement entropy with Eqs.~\eqref{eq:scale1} and~\eqref{eq:scale2} over a certain range of bond dimensions $\mathcal{I}_{\chi}$. 
By defining $\alpha$ and $\beta$ such that
\begin{align}
	S = \beta \log(\chi) \\
	\log(\psi) = \alpha \log(\chi).
\end{align}
and assuming the central charge we then get $\kappa = 6\beta/c$ and $\eta= -2\alpha/\kappa$ and the computation of the critical exponent is reduced to a simple linear fit. As the environment converges modulo three, the entanglement entropy takes three different values $S_A,S_B,S_C$ depending on the modulo of the number of iterations. Furthermore, by symmetry of the model we have $S_B = S_C$ and we can then define two different exponents $\eta_A \equiv \eta(S_A)$ and $\eta_B \equiv \eta(S_B)$ depending on which entropy $S_A$ or $S_B$ we choose to define $\kappa$ from. We benchmark the methodology at the Villain-Stephenson point at $H = K^{-1} = 0$ using bond dimension $\chi \in [200,400]$ and compare the results with the square CTMRG. The results are shown in Fig. \ref{fig:comparison}. Assuming $c =  1$, we found for the honeycomb CTMRG $\eta_A = 0.4990 \pm 0.0027$ and $\eta_B = 0.4986 \pm 0.0027$ in good agreement with the exact value $\eta = 1/2$ where the error bars have been computed from the goodness of the fit. Conversely, one could have assumed $\eta = 1/2$ and we recover $c_A = 1.002\pm 0.005$ and $c_B = 1.003\pm 0.005$. However, we note that even by assuming the central charge $c = 1$ we find $\kappa_A = 1.230\pm 0.003$ and $\kappa_B = 1.231 \pm 0.003$ which do not agree with the theoretical prediction given by Eq.~\eqref{eq:kappa}. We attribute this behavior to various corrections in bond dimension~\cite{Okunishi2022, Schmoll2019, Ueda2020}. We then choose to assume $\kappa$ as an independent parameter to be measured from the entanglement entropy rather than using the theoretical prediction. We note that we cannot measure $\kappa$ from looking at the correlation length scaling with respect to the bond dimension as the algorithm at zero temperature already spontaneously breaks translational symmetry and we lose access to the transfer matrix. Applying the same methodology with the square CTMRG and assuming the central charge $c = 1$ we found $\eta = 0.4839\pm 0.0131$. Furthermore, assuming $\eta = 1/2$ we obtain $c = 1.0334\pm0.0280$. The critical exponent and central charge obtained from the square CTMRG are less accurate than the ones obtained with the $C_3$-symmetric CTMRG with larger error bars. This is due to noise present in the fitting of the order parameter and entanglement entropy seen in Fig. \ref{fig:comparison}, and in that instance the $C_3$-symmetric CTMRG outperforms the square CTMRG.  

We also note that the exponential divergence of the correlation length limits the range of numerically reachable temperatures. Indeed, since the numerically accessible correlation length is also bound by the bond dimension with $\xi_\chi = \chi^{\kappa}$, at low enough temperature the (physical) correlation length will be larger than what can be described with finite bond dimension $\chi$, $i.e.$ $\xi > \xi_\chi$ and the system will freeze in the sense that (i) the correlation length saturates to a finite value and (ii) correspondingly, the environment converges to a temperature-independent fixed point. Due to the exponential divergence of the correlation length, in order to overcome this problem and to effectively describe the lower temperatures we would need to increase $\chi$ exponentially fast as well. 

\subsection{Constrained model}

\begin{figure}[t!]
\includegraphics[width=0.45\textwidth]{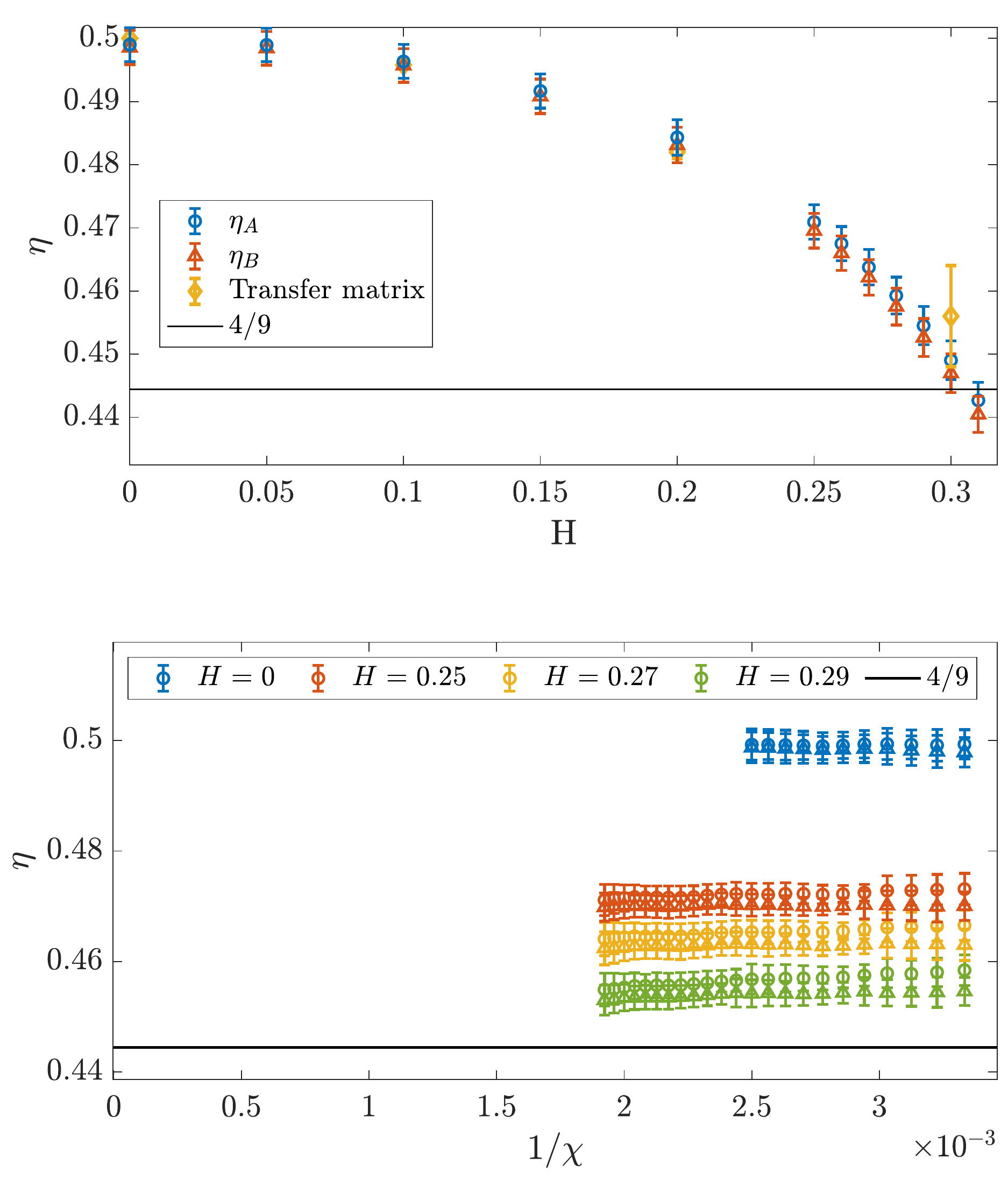}
\caption{Upper panel : Simulations done at $K^{-1}= 0$. At $H = 0$, the results are in good agreement with the exact results $\eta = 1/2$. For $H\leq 0.2,$ we used bond dimension $\chi \in [200,400]$, while for $H>0.2$ we used $\chi \in [300,520]$. The transfer matrix results are from Ref.~\onlinecite{Blote93}. Lower panel : simulations done at $K^{-1}= 0$. The $\circ$ and $\triangle$ symbols denote $\eta_A$ and $\eta_B$, respectively. In the $\chi$ goes to infinity limit one recovers the exact result.}
\label{fig:eta_dep}
\end{figure}

We now turn to the constrained model by setting $K^{-1} = 0$. As already mentioned, there are two phases separated by a KT transition characterized by $\eta_{KT} = 4/9$. We thus use the critical value of $\eta$ as the criterion to locate the transition and we found $H_{KT} = 0.305\pm0.006$. The exponent is computed with the scaling relations discussed in the previous section and the results are shown in Fig.~\ref{fig:eta_dep} (upper panel). For $H \leq 0.2$ we used bond dimension $\chi \in [200,400]$ while for $H \geq 0.25$, closer to the transition, we used $\chi \in [300,520]$. We have plotted the transfer matrix results obtained from finite-size scaling assuming power law correction (Ref.~\onlinecite{Blote93}, Table II) as well. At low reduced fields, our results agree reasonably well with the transfer matrix results while at larger reduced field, the two methods start to give different results. It is worth noting that although the values of the exponent obtained from transfer matrix above $H>0.3$ are attributed in Ref.~\onlinecite{Blote93} to strong cross-overs into the ordered phase (see also~\cite{Otsuka06,Hwang2010}), using the $\eta_{KT} = 4/9$ criteria to determine the location of the KT transition would lead to a larger value than the present result.

\par Indeed, around the transition, significant corrections to scaling are expected~\cite{Blote93} and the critical field $H_{KT} = 0.266\pm 0.01$ in Refs.~\onlinecite{Blote91,Blote93, Qian} is only obtained by considering \emph{both} finite size logarithmic and power-law corrections. It is unclear how to perform similar corrections in the framework of finite-bond dimension scaling. We note that, close to $H\sim 0.3$, our results show a slight dependence on the range of bond dimension over which we fit: when considering higher bond dimensions, the computation of $\eta$ gives a lower exponent. We illustrate the dependency in Fig.~\ref{fig:eta_dep} (bottom panel) for three different values of the reduced field where in the $x$-axis $\chi$ is used as an abuse of notation referring to the interval $\mathcal{I}_{\chi} = [\chi-250,\chi]$. As expected, the difference between $\eta_A$ and $\eta_B$ decreases upon increasing the bond dimension. At $H = 0$ and $H = 0.25$ the CTMRG results are well converged. In contrast, at $H = 0.29$ we observe some dependency and it becomes harder to conclude. From the behavior of $\eta$ in the lower-panel of Fig.~\ref{fig:eta_dep}, we draw the conclusion that the obtained critical field should be taken as an upper bound for the transition.

\subsection{High field}

\begin{figure}[t!]
\includegraphics[width=0.45\textwidth]{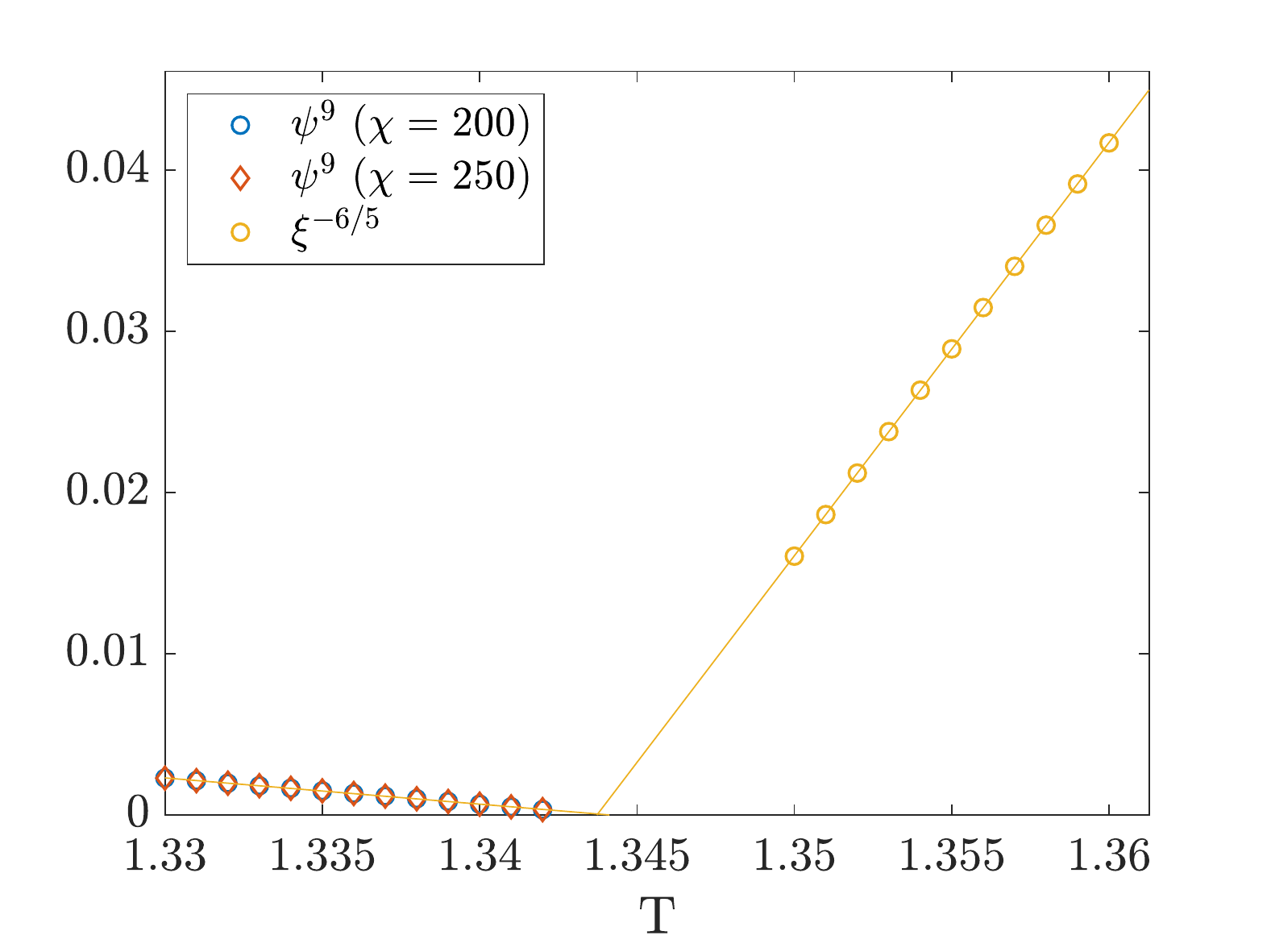}
\caption{Simulations were done at $h = 3$. The order parameter has been computed at fixed bond dimension. The critical temperatures obtained from the order parameter and correlation length respectively are $T_c = 1.3440$, and $T_c = 1.3437$.}
\label{fig:bm1}
\end{figure}

At high field, the critical temperature is large and both transfer matrices and Monte Carlo methods also have no problem identifying the location of the Potts transition or its universality class\cite{Alexander,Kinzel1981,Qian}. As an additional benchmark of our approach, we verify the nature of the transition at $h = 3$: if the transition belongs to the three-state Potts universality class, both $\psi^9$ and $\xi^{-6/5}$ should behave linearly and intersect the $x$-axis at the critical temperature. This is indeed what we observe in Fig.~\ref{fig:bm1}. More precisely, the intersection between the linear fit of $\psi^9$ and the $x$-axis gives a critical temperature $T_c = 1.3440$ while the intersection between the linear fit of $\xi^{-6/5}$ and the $x$-axis gives a critical temperature $T_c = 1.3437$. The difference between the two temperatures is of the order of $10^{-4}$, in agreement with a unique transition belonging to the three-state Potts universality class. The correlation length has been extrapolated using bond dimension ranging from $\chi = 100$ to $\chi = 200$. We have used $\chi = 200$ and $\chi = 250$ for the order parameter and we can see in Fig.~\ref{fig:bm1} that it has converged with respect to the bond dimension.

\subsection{Low field}

We now discuss in detail the results obtained at low temperatures. We observe a strong bond dimension dependency for the order parameter and entanglement entropy, even away from the critical line. Yet, as previously discussed, such dependency can be used to determine the critical temperatures and critical exponents. In contrast to the previous section, we do not work at fixed field to check the nature of the universality class, but we are looking for the location of the critical reduced field knowing that the transition is in the three-state Potts universality class. We thus choose to locate the transition by looking at the regime where the order parameter decays with respect to $\chi$ as a power law and perform a self-consistent check by assuming the central charge $c = 4/5$ and then measuring the spin-spin decay critical exponent $\eta$. 

\begin{figure}[t!]
\includegraphics[width=0.45\textwidth]{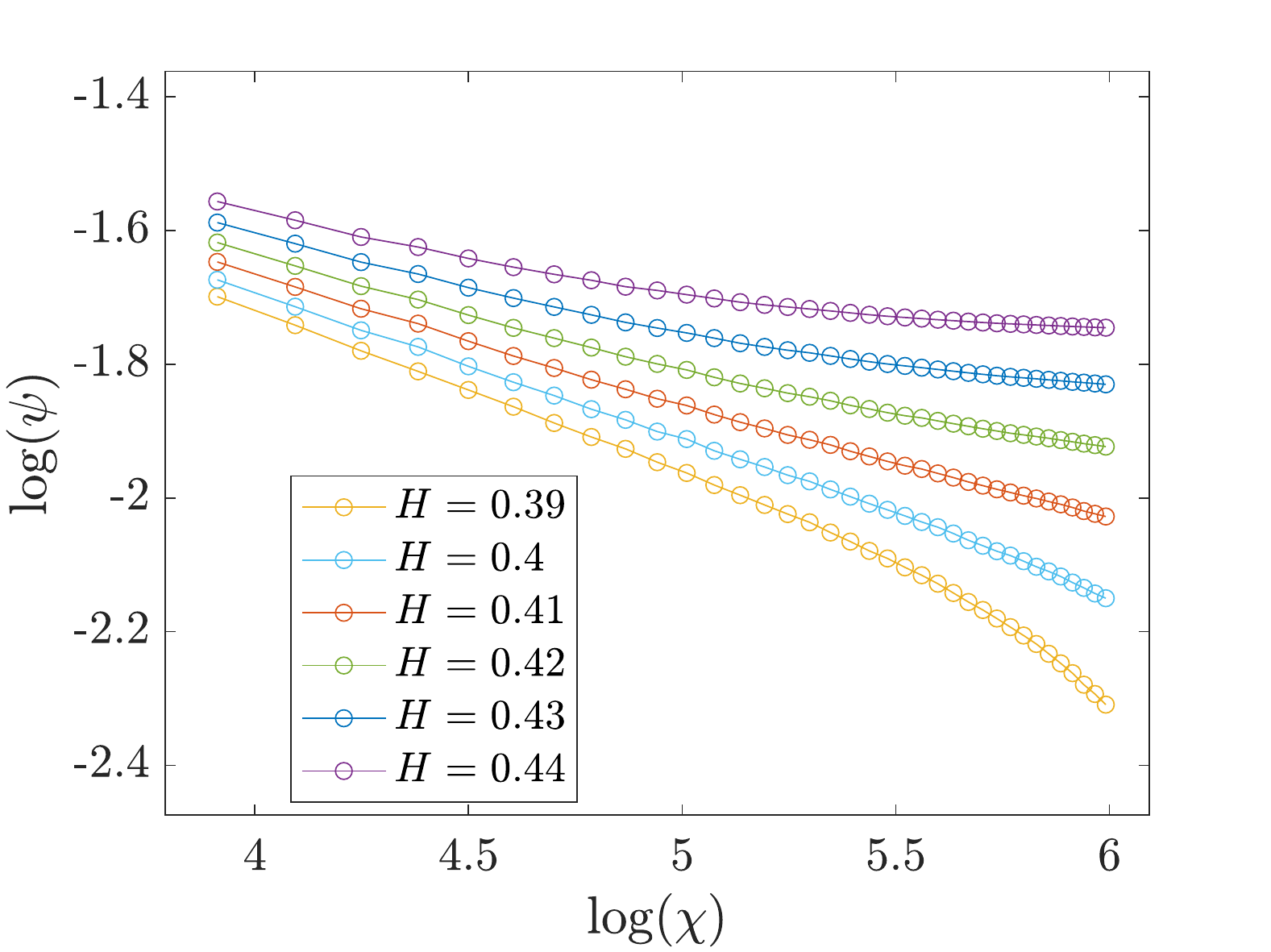}
\caption{log-log plot of the order parameter with respect to the bond dimension at $K^{-1} = 0.25$. We can identify the power law regime to be in between $H = 0.40$ and $H = 0.42$. We thus estimate $H_c = 0.41 \pm 0.01$. At $H_c = 0.41$, by fitting the order parameter for $\chi \in [200,400]$ one finds $\eta_A = 0.270\pm0.003$ and $\eta_B = 0.266 \pm 0.003$ in agreement with the three-state Potts universality class.}
\label{fig:ChiTc}
\end{figure}

We show the results in Fig.~\ref{fig:ChiTc} for the lowest considered temperature $K^{-1} = 0.25$ where we used bond dimension up to $\chi = 400$. The algebraic decay of the order parameter can be seen in only a narrow interval $H_c \in [0.40, 0.42]$. By considering higher bond dimensions, the interval's width would diminish. But we are limited to finite bond dimension and thus consider $H_c = 0.41 \pm 0.01$. Assuming $c = 4/5$ and considering only the largest bond dimensions between $\chi = 200$ and $\chi = 400$ we find $\eta_A = 0.270\pm0.003$ and $\eta_B = 0.266 \pm 0.003$ in good agreement with the three-state Potts universality class. By recovering the right critical exponent we confirm that we are not in a cross-over regime influenced by the proximity of the KT point and that we indeed have located the critical reduced field. 

\begin{figure}[t!]
\includegraphics[width=0.45\textwidth]{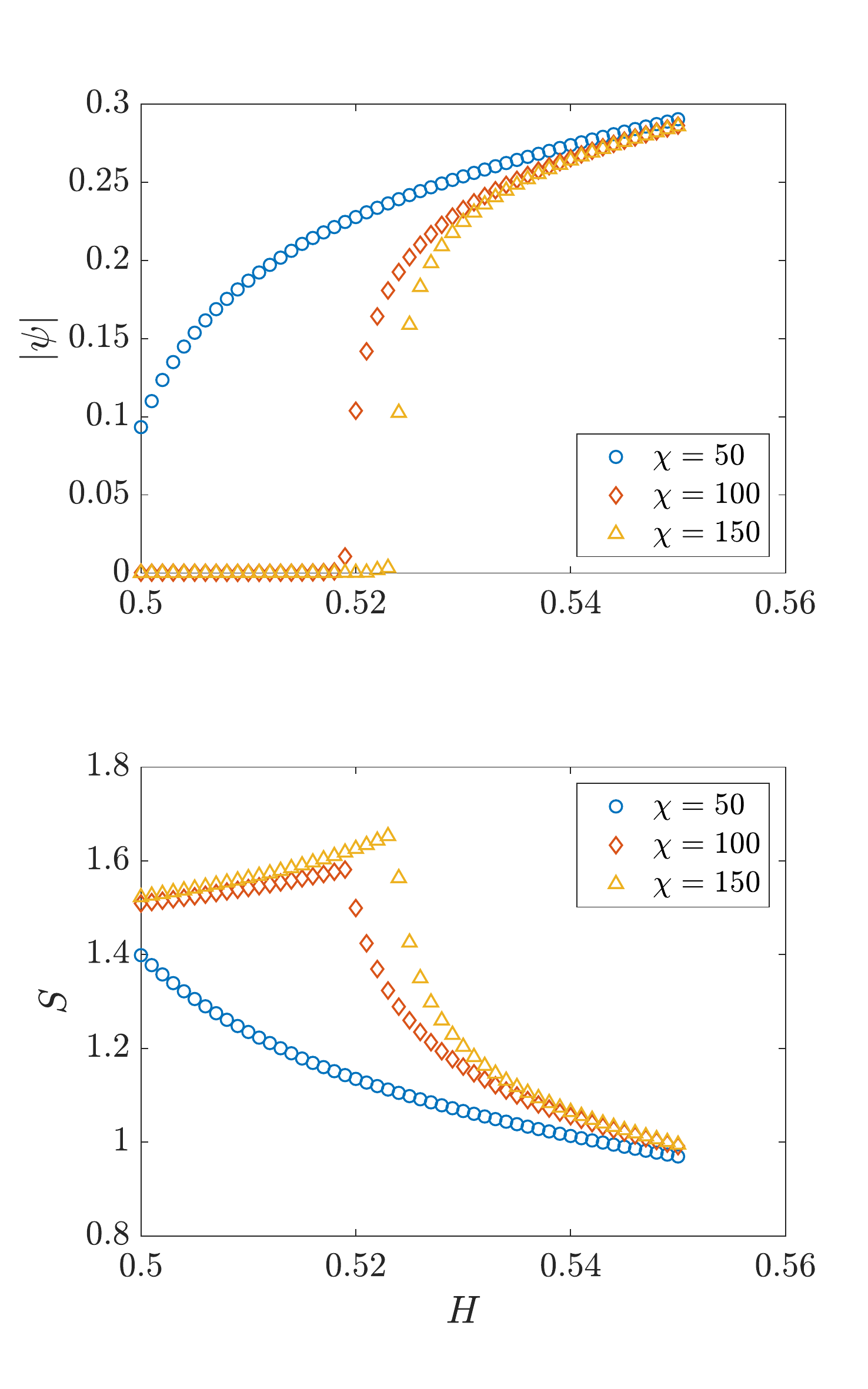}
\caption{Order parameter and entanglement entropy at temperature $K^{-1} = 0.4$ for different bond dimensions.}
\label{fig:chidatas}
\end{figure}

\subsection{Phase Diagram}

\begin{figure}[t!]
\includegraphics[width=0.45\textwidth]{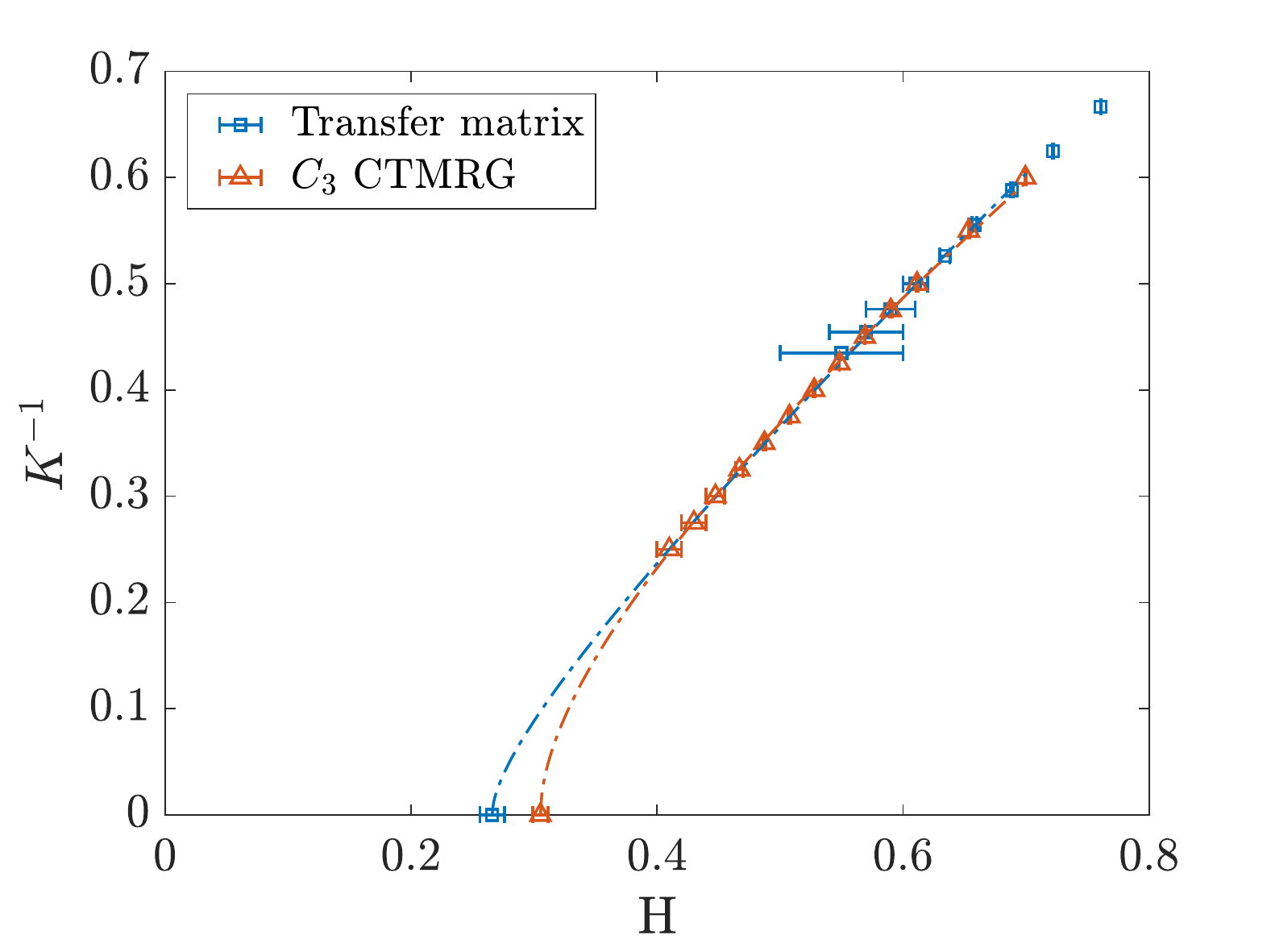}
\caption{Phase diagram of the triangular antiferromagnetic Ising model in a field shown as $K^{-1}$ with respect to the reduced field $H$. Our results are in good agreement with previous transfer matrix study. Blue and red dashed lines indicate the fits done up to second order on the present transition considering respectively  $H_{KT} = 0.266\pm 0.01$ and $H_{KT} = 0.305\pm 0.006$.}
\label{fig:phase_diagram}
\end{figure}

The full phase diagram is given in Fig.~\ref{fig:phase_diagram}, where we have plotted the critical temperature obtained from the $C_3$ symmetric honeycomb CTMRG as well as the transfer matrix results obtained in Ref.~\onlinecite{Qian}. At finite temperature, our results agree reasonably well with the data from the transfer matrix methods, but with significantly smaller error bars. Furthermore, we are able to reach lower temperatures while keeping reasonably small errorbars, such that the transition line we found shows the first signs of a curvature, which could not be observed previously. In that regard, our study further supports the renormalization group predictions which suggest that the Potts critical line meets the zero temperature KT point with a leading exponent one-half and a linear correction~\cite{Nienhuis_1984, Qian}: 
\begin{equation}
K_c^{-1}\propto (H_c - H_{KT} )^{1/2} + o(H_c - H_{KT}).
\end{equation}

We have also fitted the new transition line up to the second order in that expansion and show the results in Fig.~\ref{fig:phase_diagram}. We find that the fit of the transition line is perfectly compatible with both critical reduced fields $H_{KT} = 0.266\pm 0.01$ and $H_{KT} = 0.305\pm 0.006$ and is not enough to distinguish between the two.

As previously mentioned, due to the exponential divergence of the correlation length upon approaching the $K^{-1} = 0$ critical line, there is a limitation for the temperatures we could reach, and we were not able to study the phase diagram for non zero temperature smaller than $K^{-1} < 0.25$.  

\section{Discussion and Summary}
\label{sec:Discussion}

We have introduced a variant of the CTMRG algorithm that contracts infinite honeycomb tensor networks. It is especially powerful for problems that have a natural $C_3$-symmetry. In the case of the antiferromagnetic triangular Ising model, this method is more accurate than the square CTMRG and when considering an extra field outperforms the more standard transfer matrix methods.
By exploring the triangular Ising antiferromagnet in a longitudinal field, we confirmed that the tensor network construction based on ground-state rules~\cite{Vanhecke2021} also allows to reach low temperatures in the honeycomb lattice CTMRG. Our results mostly confirm and support the predictions from the literature~\cite{Nienhuis_1984,Blote93, Qian}: at finite temperature, as expected, we found the nature of the transition to be three-state Potts by measuring a unique critical temperature from the correlation length and order parameter but also by using bond dimension scaling of the entanglement entropy and order parameter. Overall, we confirm Qian \textit{et al.} results by recovering essentially the same critical temperatures but with significantly smaller error bars, and we extend the transition line to lower temperatures previously not available. As the temperature is lowered to zero, we found evidence of the reduced critical field converging to the Kosterlitz-Thouless critical reduced field of the constrained model in favor of the scenario proposed by Nienhuis \textit{et al.}. At $T = 0$, we locate the KT transition at a slightly higher reduced field than previously reported.

Several improvements of this CTMRG for the honeycomb lattice might be worth investigating. The choice of isometry could be different: the most commonly used isometry in the square lattice CTMRG is the one first introduced by Corboz for contracting iPEPS~\cite{IsoCorboz}. It would be interesting to formulate the equivalent isometry for CTMRG on honeycomb and to see whether it leads to a better convergence rate for some problems. A more challenging improvement would be to design a more general multisite version of the algorithm which can take arbitrary unit cell; as mentioned, this could be useful to have access to the transfer matrix also in presence of symmetry breaking. Furthermore, this algorithm can also be used to contract two-dimensional quantum systems iPEPS wave-functions on the honeycomb lattice~\cite{Corboz2012Spin} and can easily be combined with simple or full updates. It could also be implemented with automatic differentiation, in a similar spirit as Ref.~\onlinecite{Lukin2023}. In the framework of PESS or iPESO, the algorithm could also be used to investigate 2D quantum systems on the kagome or triangular lattices \cite{Xie2014,Schmoll2022finite,Jahromi2020,Liao2021}.  

\vskip2.cm

{\it Acknowledgments.}
We thank Loic Herviou, Olivier Gauthé, Mithilesh Nayak, Andrej Gendiar and Bram Vanhecke for useful discussions. 
This work has been supported by the Swiss National Science Foundation grant No. 212082.
The calculations have been performed using the facilities of the Scientific IT and Application Support Center of EPFL.

\bibliography{bibliography}

\end{document}